\documentclass[draftcls,onecolumn, 12pt]{IEEEtran}

\makeatletter
\def\endthebibliography{%
	\def\@noitemerr{\@latex@warning{Empty `thebibliography' environment}}%
	\endlist
}
\makeatother

\usepackage{silence}
\usepackage{verbatim}
\usepackage{xr}

\PassOptionsToPackage{hyphens}{url}

\usepackage[T1]{fontenc}
\usepackage[utf8]{inputenc}
\usepackage{mathtools}

\usepackage{amssymb,mathrsfs}
\usepackage{amsthm}
\usepackage{bm}
\usepackage{scalerel}
\usepackage{nicefrac}
\usepackage{microtype} 
\usepackage[shortlabels]{enumitem}
\usepackage{graphicx}
\usepackage{epstopdf}
\DeclareGraphicsExtensions{.eps,.png,.jpg,.pdf}

\usepackage{url}
\usepackage{colortbl}
\usepackage{booktabs}
\usepackage{multirow}
\usepackage{colortbl,xcolor}
\usepackage[normalem]{ulem}
\usepackage{xparse,xstring}
\usepackage{calc}
\usepackage{etoolbox}

\makeatletter
\@ifpackageloaded{natbib}{
	\relax
}{
	\usepackage{cite}
}
\makeatother

\usepackage{array}
\newcolumntype{L}[1]{>{\raggedright\let\newline\\\arraybackslash\hspace{0pt}}m{#1}}
\newcolumntype{C}[1]{>{\centering\let\newline\\\arraybackslash\hspace{0pt}}m{#1}}
\newcolumntype{R}[1]{>{\raggedleft\let\newline\\\arraybackslash\hspace{0pt}}m{#1}}

\makeatletter
\let\MYcaption\@makecaption
\makeatother
\usepackage[font=footnotesize]{subcaption}
\makeatletter
\let\@makecaption\MYcaption
\makeatother

\usepackage{glossaries}
\makeatletter
\sfcode`\.1006

\let\oldgls\gls
\let\oldglspl\glspl

\newcommand\fussy@ifnextchar[3]{%
	\let\reserved@d=#1%
	\def\reserved@a{#2}%
	\def\reserved@b{#3}%
	\futurelet\@let@token\fussy@ifnch}
\def\fussy@ifnch{%
	\ifx\@let@token\reserved@d
		\let\reserved@c\reserved@a
	\else
		\let\reserved@c\reserved@b
	\fi
	\reserved@c}

\renewcommand{\gls}[1]{%
\oldgls{#1}\fussy@ifnextchar.{\@checkperiod}{\@}}
\renewcommand{\glspl}[1]{%
\oldglspl{#1}\fussy@ifnextchar.{\@checkperiod}{\@}}

\newcommand{\@checkperiod}[1]{%
	\ifnum\sfcode`\.=\spacefactor\else#1\fi
}

\robustify{\gls}
\robustify{\glspl}
\makeatother

\newacronym{wrt}{w.r.t.}{with respect to}
\newacronym{RHS}{R.H.S.}{right-hand side}
\newacronym{LHS}{L.H.S.}{left-hand side}
\newacronym{iid}{i.i.d.}{independent and identically distributed}
\newacronym{SOTA}{SOTA}{state-of-the-art}

\usepackage{float}

\ifx\notloadhyperref\undefined
	\ifx\loadbibentry\undefined
		\usepackage[hidelinks,hypertexnames=false]{hyperref}
	\else
		\usepackage{bibentry}
		\makeatletter\let\saved@bibitem\@bibitem\makeatother
		\usepackage[hidelinks,hypertexnames=false]{hyperref}
		\makeatletter\let\@bibitem\saved@bibitem\makeatother
	\fi
\else
	\ifx\loadbibentry\undefined
		\relax
	\else
		\usepackage{bibentry}
	\fi
\fi

\usepackage[capitalize]{cleveref}
\crefname{equation}{}{}
\Crefname{equation}{}{}
\crefname{claim}{claim}{claims}
\crefname{step}{step}{steps}
\crefname{line}{line}{lines}
\crefname{condition}{condition}{conditions}
\crefname{dmath}{}{}
\crefname{dseries}{}{}
\crefname{dgroup}{}{}

\crefname{Problem}{Problem}{Problems}
\crefformat{Problem}{Problem~#2#1#3}
\crefrangeformat{Problem}{Problems~#3#1#4 to~#5#2#6}

\crefname{Theorem}{Theorem}{Theorems}
\crefname{Corollary}{Corollary}{Corollaries}
\crefname{Proposition}{Proposition}{Propositions}
\crefname{Lemma}{Lemma}{Lemmas}
\crefname{Definition}{Definition}{Definitions}
\crefname{Example}{Example}{Examples}
\crefname{Assumption}{Assumption}{Assumptions}
\crefname{Remark}{Remark}{Remarks}
\crefname{Rem}{Remark}{Remarks}
\crefname{remarks}{Remarks}{Remarks}
\crefname{Appendix}{Appendix}{Appendices}
\crefname{Supplement}{Supplement}{Supplements}
\crefname{Exercise}{Exercise}{Exercises}
\crefname{Theorem_A}{Theorem}{Theorems}
\crefname{Corollary_A}{Corollary}{Corollaries}
\crefname{Proposition_A}{Proposition}{Propositions}
\crefname{Lemma_A}{Lemma}{Lemmas}
\crefname{Definition_A}{Definition}{Definitions}

\usepackage{crossreftools}
\ifx\notloadhyperref\undefined
\else
	\relax
\fi

\usepackage{algorithm}
\usepackage{algpseudocode}

\ifx\loadbreqn\undefined
	\relax
\else
	\usepackage{breqn}
\fi

\makeatletter
\def\cleartheorem#1{%
    \expandafter\let\csname#1\endcsname\relax
    \expandafter\let\csname c@#1\endcsname\relax
}
\def\clearthms#1{ \@for\tname:=#1\do{\cleartheorem\tname} }
\makeatother

\ifx\renewtheorem\undefined
	\ifx\useTheoremCounter\undefined
		\newtheorem{Theorem}{Theorem}
		\newtheorem{Corollary}{Corollary}
		\newtheorem{Proposition}{Proposition}
		\newtheorem{Lemma}{Lemma}
	\else
		\newtheorem{Theorem}{Theorem}

	\fi

	\newtheorem{Definition}{Definition}
	\newtheorem{Example}{Example}
	
	\newtheorem{Assumption}{Assumption}

\fi

\theoremstyle{remark}

\theoremstyle{plain}

\newcommand{\qednew}{\nobreak \ifvmode \relax \else
		\ifdim\lastskip<1.5em \hskip-\lastskip
			\hskip1.5em plus0em minus0.5em \fi \nobreak
		\vrule height0.75em width0.5em depth0.25em\fi}



\newcommand{\nn}{\nonumber\\ }

\NewDocumentCommand{\movedownsub}{e{^_}}{%
	\IfNoValueTF{#1}{%
		\IfNoValueF{#2}{^{}}
	}{%
		^{#1}
	}%
	\IfNoValueF{#2}{_{#2}}
}

\let\latexchi\chi
\RenewDocumentCommand{\chi}{}{\latexchi\movedownsub}

\newcommand{\Real}{\mathbb{R}}
\newcommand{\Nat}{\mathbb{N}}

\newcommand{\calA}{\mathcal{A}}
\newcommand{\calB}{\mathcal{B}}
\newcommand{\calC}{\mathcal{C}}
\newcommand{\calD}{\mathcal{D}}
\newcommand{\calE}{\mathcal{E}}
\newcommand{\calF}{\mathcal{F}}
\newcommand{\calG}{\mathcal{G}}
\newcommand{\calH}{\mathcal{H}}
\newcommand{\calI}{\mathcal{I}}
\newcommand{\calJ}{\mathcal{J}}
\newcommand{\calK}{\mathcal{K}}

\newcommand{\calN}{\mathcal{N}}

\newcommand{\calP}{\mathcal{P}}
\newcommand{\calQ}{\mathcal{Q}}

\newcommand{\calS}{\mathcal{S}}
\newcommand{\calT}{\mathcal{T}}

\newcommand{\calV}{\mathcal{V}}

\newcommand{\calX}{\mathcal{X}}
\newcommand{\calY}{\mathcal{Y}}
\newcommand{\calZ}{\mathcal{Z}}

\newcommand{\bA}{\mathbf{A}}

\newcommand{\bB}{\mathbf{B}}
\newcommand{\bc}{\mathbf{c}}
\newcommand{\bC}{\mathbf{C}}

\newcommand{\bD}{\mathbf{D}}
\newcommand{\be}{\mathbf{e}}

\newcommand{\boldf}{\mathbf{f}}

\newcommand{\bh}{\mathbf{h}}
\newcommand{\bH}{\mathbf{H}}

\newcommand{\bI}{\mathbf{I}}

\newcommand{\bJ}{\mathbf{J}}
\newcommand{\bk}{\mathbf{k}}
\newcommand{\bK}{\mathbf{K}}

\newcommand{\bL}{\mathbf{L}}
\newcommand{\boldm}{\mathbf{m}}

\newcommand{\bp}{\mathbf{p}}

\newcommand{\bQ}{\mathbf{Q}}

\newcommand{\bs}{\mathbf{s}}
\newcommand{\bS}{\mathbf{S}}
\newcommand{\bt}{\mathbf{t}}

\newcommand{\bv}{\mathbf{v}}

\newcommand{\bw}{\mathbf{w}}

\newcommand{\bx}{\mathbf{x}}
\newcommand{\bX}{\mathbf{X}}
\newcommand{\by}{\mathbf{y}}
\newcommand{\bY}{\mathbf{Y}}
\newcommand{\bz}{\mathbf{z}}

\newcommand{\bbC}{\mathbb{C}}

\newcommand{\bbN}{\mathbb{N}}

\DeclareSymbolFont{bsfletters}{OT1}{cmss}{bx}{n}
\DeclareSymbolFont{ssfletters}{OT1}{cmss}{m}{n}
\DeclareMathSymbol{\bsfGamma}{0}{bsfletters}{'000}
\DeclareMathSymbol{\ssfGamma}{0}{ssfletters}{'000}
\DeclareMathSymbol{\bsfDelta}{0}{bsfletters}{'001}
\DeclareMathSymbol{\ssfDelta}{0}{ssfletters}{'001}
\DeclareMathSymbol{\bsfTheta}{0}{bsfletters}{'002}
\DeclareMathSymbol{\ssfTheta}{0}{ssfletters}{'002}
\DeclareMathSymbol{\bsfLambda}{0}{bsfletters}{'003}
\DeclareMathSymbol{\ssfLambda}{0}{ssfletters}{'003}
\DeclareMathSymbol{\bsfXi}{0}{bsfletters}{'004}
\DeclareMathSymbol{\ssfXi}{0}{ssfletters}{'004}
\DeclareMathSymbol{\bsfPi}{0}{bsfletters}{'005}
\DeclareMathSymbol{\ssfPi}{0}{ssfletters}{'005}
\DeclareMathSymbol{\bsfSigma}{0}{bsfletters}{'006}
\DeclareMathSymbol{\ssfSigma}{0}{ssfletters}{'006}
\DeclareMathSymbol{\bsfUpsilon}{0}{bsfletters}{'007}
\DeclareMathSymbol{\ssfUpsilon}{0}{ssfletters}{'007}
\DeclareMathSymbol{\bsfPhi}{0}{bsfletters}{'010}
\DeclareMathSymbol{\ssfPhi}{0}{ssfletters}{'010}
\DeclareMathSymbol{\bsfPsi}{0}{bsfletters}{'011}
\DeclareMathSymbol{\ssfPsi}{0}{ssfletters}{'011}
\DeclareMathSymbol{\bsfOmega}{0}{bsfletters}{'012}
\DeclareMathSymbol{\ssfOmega}{0}{ssfletters}{'012}

\newcommand{\bmeta}{\bm{\eta}}
\newcommand{\bkappa}{\bm{\kappa}}

\newcommand{\bphi}{\bm{\phi}}
\newcommand{\bpsi}{\bm{\psi}}

\newcommand{\bxi}{\bm{\xi}}

\newcommand{\bLambda}{\bm{\Lambda}}

\newcommand{\bPhi}{\bm{\Phi}}
\newcommand{\bPi}{\bm{\Pi}}

\makeatletter
\newcommand*\rel@kern[1]{\kern#1\dimexpr\macc@kerna}
\newcommand*\widebar[1]{%
  \begingroup
  \def\mathaccent##1##2{%
    \rel@kern{0.8}%
    \overline{\rel@kern{-0.8}\macc@nucleus\rel@kern{0.2}}%
    \rel@kern{-0.2}%
  }%
  \macc@depth\@ne
  \let\math@bgroup\@empty \let\math@egroup\macc@set@skewchar
  \mathsurround\z@ \frozen@everymath{\mathgroup\macc@group\relax}%
  \macc@set@skewchar\relax
  \let\mathaccentV\macc@nested@a
  \macc@nested@a\relax111{#1}%
  \endgroup
}
\makeatother

\DeclareMathOperator*{\argmin}{arg\,min}

\DeclareMathOperator{\ST}{s.t.\ }
\DeclareMathOperator{\as}{a.s.}
\DeclareMathOperator{\const}{const}
\DeclareMathOperator{\diag}{diag}

\DeclareMathOperator{\tr}{tr}

\DeclareMathOperator{\spn}{span}

\DeclareMathOperator{\var}{var}

\DeclareMathOperator{\cov}{cov}
\DeclareMathOperator{\corr}{corr}

\DeclareMathOperator{\vect}{vec}

\DeclareMathOperator{\ima}{im}
\DeclareMathOperator{\Mod}{mod}

\newcommand{\ifbcdot}[1]{\ifblank{#1}{\cdot}{#1}}

\DeclarePairedDelimiterX\abs[1]{\lvert}{\rvert}{\ifbcdot{#1}}
\DeclarePairedDelimiterX\parens[1]{(}{)}{\ifbcdot{#1}}
\DeclarePairedDelimiterX\brk[1]{[}{]}{\ifbcdot{#1}}
\DeclarePairedDelimiterX\braces[1]{\{}{\}}{\ifbcdot{#1}}
\DeclarePairedDelimiterX\angles[1]{\langle}{\rangle}{\ifblank{#1}{\cdot,\cdot}{#1}}
\DeclarePairedDelimiterX\ip[2]{\langle}{\rangle}{\ifbcdot{#1},\ifbcdot{#2}}
\DeclarePairedDelimiterX\norm[1]{\lVert}{\rVert}{\ifbcdot{#1}}
\DeclarePairedDelimiterX\ceil[1]{\lceil}{\rceil}{\ifbcdot{#1}}
\DeclarePairedDelimiterX\floor[1]{\lfloor}{\rfloor}{\ifbcdot{#1}}

\DeclareFontFamily{U}{matha}{\hyphenchar\font45}
\DeclareFontShape{U}{matha}{m}{n}{
      <5> <6> <7> <8> <9> <10> gen * matha
      <10.95> matha10 <12> <14.4> <17.28> <20.74> <24.88> matha12
      }{}
\DeclareSymbolFont{matha}{U}{matha}{m}{n}
\DeclareFontSubstitution{U}{matha}{m}{n}

\DeclareFontFamily{U}{mathx}{\hyphenchar\font45}
\DeclareFontShape{U}{mathx}{m}{n}{
      <5> <6> <7> <8> <9> <10>
      <10.95> <12> <14.4> <17.28> <20.74> <24.88>
      mathx10
      }{}
\DeclareSymbolFont{mathx}{U}{mathx}{m}{n}
\DeclareFontSubstitution{U}{mathx}{m}{n}

\DeclareMathDelimiter{\vvvert}{0}{matha}{"7E}{mathx}{"17}
\DeclarePairedDelimiterX\vertiii[1]{\vvvert}{\vvvert}{\ifbcdot{#1}}

\DeclarePairedDelimiterXPP\trace[1]{\operatorname{Tr}}{(}{)}{}{\ifbcdot{#1}} 
\DeclarePairedDelimiterXPP\col[1]{\operatorname{col}}{\{}{\}}{}{\ifbcdot{#1}} 
\DeclarePairedDelimiterXPP\row[1]{\operatorname{row}}{\{}{\}}{}{\ifbcdot{#1}} 
\DeclarePairedDelimiterXPP\erf[1]{\operatorname{erf}}{(}{)}{}{\ifbcdot{#1}}
\DeclarePairedDelimiterXPP\erfc[1]{\operatorname{erfc}}{(}{)}{}{\ifbcdot{#1}}
\DeclarePairedDelimiterXPP\KLD[2]{D}{(}{)}{}{\ifbcdot{#1}\, \delimsize\|\, \ifbcdot{#2}} 
\DeclarePairedDelimiterXPP\op[2]{\operatorname{#1}}{(}{)}{}{#2} 

\newcommand{\T}{^{\mkern-1.5mu\mathop\intercal}}
\newcommand{\setcomp}{^{\mathsf{c}}} 
\newcommand{\ud}{\,\mathrm{d}} 

\newcommand{\bone}{\bm{1}}

\DeclarePairedDelimiterXPP\indicate[1]{{\bf 1}}{\{}{\}}{}{\ifbcdot{#1}}

\newcommand{\tc}[1]{^{(#1)}}
\NewDocumentCommand\ofrac{s m}{%
	\IfBooleanTF#1%
	{\dfrac{1}{#2}}%
	{\frac{1}{#2}}%
}
\NewDocumentCommand\ddfrac{s m m}{%
	\IfBooleanTF#1%
	{\dfrac{\mathrm{d} {#2}}{\mathrm{d} {#3}}}%
	{\frac{\mathrm{d} {#2}}{\mathrm{d} {#3}}}%
}
\NewDocumentCommand\ppfrac{s m m}{%
	\IfBooleanTF#1%
	{\dfrac{\partial {#2}}{\partial {#3}}}%
	{\frac{\partial {#2}}{\partial {#3}}}%
}

\providecommand\given{}

\DeclarePairedDelimiterX\Set[2]\{\}{%
\renewcommand\given{\SetSymbol[\delimsize]{#1}}
#2
}
\DeclarePairedDelimiterX\Setc[1]\{\}{%
\renewcommand\given{\SetSymbol{:}}
#1
}

\NewDocumentCommand\set{s o m}{%
	\IfBooleanTF#1%
	{\IfValueTF{#2}{\Set*{#2}{#3}}{\Setc*{#3}}}%
	{\IfValueTF{#2}{\Set{#2}{#3}}{\Setc{#3}}}%
}

\NewDocumentCommand{\evalat}{ s O{\big} m e{_^} }{%
\IfBooleanTF{#1}%
{\left. #3 \right|}{#3#2|}%
\IfValueT{#4}{_{#4}}%
\IfValueT{#5}{^{#5}}%
}

\providecommand\given{}
\DeclarePairedDelimiterXPP\cprob[1]{}(){}{
\renewcommand\given{\nonscript\,\delimsize\vert\allowbreak\nonscript\,\mathopen{}}%
#1%
}
\DeclarePairedDelimiterXPP\cexp[1]{}[]{}{
\renewcommand\given{\nonscript\,\delimsize\vert\allowbreak\nonscript\,\mathopen{}}%
#1%
}

\DeclareDocumentCommand \P { s e{_^} d() g } {%
	\mathbb{P}%
	\IfBooleanTF{#1}%
		{
			\IfValueT{#2}{_{#2}}%
			\IfValueT{#3}{^{#3}}%
			\IfValueTF{#5}{\cprob{#4 \given #5}}{\IfValueT{#4}{\cprob{#4}}}%
		}%
		{
			\IfValueT{#2}{_{#2}}%
			\IfValueT{#3}{^{#3}}%
			\IfValueTF{#5}{\cprob*{#4 \given #5}}{\IfValueT{#4}{\cprob*{#4}}}%
		}%
}

\DeclareDocumentCommand \E { s e{_^} o g } {%
	\mathbb{E}%
	\IfBooleanTF{#1}%
		{
			\IfValueT{#2}{_{#2}}%
			\IfValueT{#3}{^{#3}}%
			\IfValueTF{#5}{\cexp{#4 \given #5}}{\IfValueT{#4}{\cexp{#4}}}%
		}%
		{
			\IfValueT{#2}{_{#2}}%
			\IfValueT{#3}{^{#3}}%
			\IfValueTF{#5}{\cexp*{#4 \given #5}}{\IfValueT{#4}{\cexp*{#4}}}%
		}%
}

\DeclareDocumentCommand \Var { s e{_^} d() g } {%
	\var%
	\IfBooleanTF{#1}%
		{
			\IfValueT{#2}{_{#2}}%
			\IfValueT{#3}{^{#3}}%
			\IfValueTF{#5}{\cprob{#4 \given #5}}{\IfValueT{#4}{\cprob{#4}}}%
		}%
		{
			\IfValueT{#2}{_{#2}}%
			\IfValueT{#3}{^{#3}}%
			\IfValueTF{#5}{\cprob*{#4 \given #5}}{\IfValueT{#4}{\cprob*{#4}}}%
		}%
}

\DeclareDocumentCommand \Cov { s e{_^} d() g } {%
	\cov%
	\IfBooleanTF{#1}%
		{
			\IfValueT{#2}{_{#2}}%
			\IfValueT{#3}{^{#3}}%
			\IfValueTF{#5}{\cprob{#4 \given #5}}{\IfValueT{#4}{\cprob{#4}}}%
		}%
		{
			\IfValueT{#2}{_{#2}}%
			\IfValueT{#3}{^{#3}}%
			\IfValueTF{#5}{\cprob*{#4 \given #5}}{\IfValueT{#4}{\cprob*{#4}}}%
		}%
}

\ExplSyntaxOn
\NewDocumentCommand \dist {m o o} {%
\mathrm{#1}\left(%
	\IfValueT{#3}{%
		\tl_if_blank:nTF{ #3 }{\cdot\, \middle|\, }{#3\, \middle|\, }%
	}
	\IfValueT{#2}{#2}%
\right)%
}
\ExplSyntaxOff

\NewDocumentCommand {\cbrace} {t+ D[]{black} D(){\widthof{#5}} m m } {%
	\begingroup%
		\color{#2}
		\IfBooleanTF{#1}{%
			\overbrace{#4}^%
		}{
			\underbrace{#4}_%
		}%
		{\parbox[c]{#3}{\centering\footnotesize{#5}}}%
	\endgroup%
}

\let\oldforall\forall
\renewcommand{\forall}{\oldforall \, }

\let\oldexist\exists
\renewcommand{\exists}{\oldexist \, }

\makeatletter

\newcommand{\rankcolor}[2]{%
	\expandafter\renewcommand\csname #1\endcsname[1]{%
		\ifblank{##1}{%
			{\color{#2} \textbf{#2}}%
		}{%
			\ifmmode
				\textcolor{#2}{\bm{##1}}%
			\else%
				{\color{#2} \textbf{##1}}%
			\fi	
		}%
	}
}

\rankcolor{first}{red}
\rankcolor{second}{blue}
\rankcolor{third}{cyan}
\makeatother

\graphicspath{{./Figures/}{./figures/}}
\DeclareDocumentCommand{\includeCroppedPdf}{ o O{./Figures/} m }{
	\IfFileExists{#2#3-crop.pdf}{}{%
		\immediate\write18{pdfcrop #2#3.pdf #2#3-crop.pdf}}%
	\includegraphics[#1]{#2#3-crop.pdf}
}

\makeatletter
\newcommand*{\addFileDependency}[1]{
  \typeout{(#1)}
  \@addtofilelist{#1}
  \IfFileExists{#1}{}{\typeout{No file #1.}}
}
\makeatother

\newcommand*{\myexternaldocument}[1]{%
    \externaldocument{#1}%
    \addFileDependency{#1.tex}%
    \addFileDependency{#1.aux}%
}

\definecolor{gray90}{gray}{0.9}
\def\colorlist{red,blue,brown,cyan,darkgray,gray,lightgray,green,lime,magenta,olive,orange,pink,purple,teal,violet,white,yellow}

\makeatletter
\def\startcomment{[}
\ifx\nohighlights\undefined
	\newcommand{\createcolor}[1]{%
			\expandafter\newcommand\csname #1\endcsname[1]{{\color{#1} ##1}}%
	}
	\newcommand{\msout}[1]{\text{\color{green} \sout{\ensuremath{#1}}}}
	\newcommand{\del}[1]{{\color{green}\ifmmode \msout{#1}\else\sout{#1}\fi}}
\else
	\newcommand{\createcolor}[1]{%
			\expandafter\newcommand\csname #1\endcsname[1]{%
				\noexpandarg%
				\StrChar{##1}{1}[\firstletter]%
				\if\firstletter\startcomment%
					\relax
				\else%
					##1
				\fi
			}%
	}
	\newcommand{\msout}[1]{}
	\newcommand{\del}[1]{}
\fi

\def\@tempa#1,{%
    \ifx\relax#1\relax\else
        \createcolor{#1}%
        \expandafter\@tempa
    \fi
}
\expandafter\@tempa\colorlist,\relax,
\makeatother

\newcommand{\hhide}[1]{}

\ifx\diagnoselabel\undefined
	\relax
\else
	\makeatletter
	\def\@testdef #1#2#3{%
		\def\reserved@a{#3}\expandafter \ifx \csname #1@#2\endcsname
			\reserved@a  \else
			\typeout{^^Jlabel #2 changed:^^J%
				\meaning\reserved@a^^J%
				\expandafter\meaning\csname #1@#2\endcsname^^J}%
			\@tempswatrue \fi}
	\makeatother
\fi

\myexternaldocument{supplementary}

\newcommand{\RH}{\Real^N\otimes\mathcal{H}}
\newcommand{\GP}{\calG\calP}
\newcommand{\iid}{\stackrel{\text{i.i.d.}}{\sim}}
\newcommand{\tens}{\circledast}

\newacronym{GSO}{GSO}{graph shift operator}
\newacronym{GSP}{GSP}{graph signal processing}
\newacronym{GFT}{GFT}{graph Fourier transform}
\newacronym{JFT}{JFT}{joint Fourier transform}
\newacronym{RKHS}{RKHS}{reproducing kernel Hilbert space}
\newacronym{JWSS}{JWSS}{joint wide-sense stationary}
\newacronym{MKL}{MKL}{multi-kernel learning}
\newacronym{WSS}{WSS}{wide-sense stationary}
\newacronym{PSD}{PSD}{power spectral density}
\newacronym{GWSS}{GWSS}{graph wide sense stationarity}
\newacronym{MSE}{MSE}{mean-squared error}
\newacronym{KRR}{KRR}{kernel ridge regression}
\newacronym{GGSP}{GGSP}{generalized graph signal processing}
\newacronym{KRG}{KRG}{kernel regression over graphs}
\newacronym{KRLS}{KRLS}{kernel recursive least-squares}
\newacronym{GP}{GP}{Gaussian process}
\newacronym{GPG}{GPG}{Gaussian process over a graph}
\newacronym{MAP}{MAP}{maximum a posteriori}
\newacronym{GRP}{GRP}{graph random process}
\newacronym{RBF}{RBF}{radial basis function}
\newacronym{SGD}{SGD}{stochastic gradient descent}
\newacronym[plural=RFFs,firstplural=random Fourier features (RFFs)]{RFF}{RFF}{random Fourier feature}
\newacronym{GTRSS}{GTRSS}{graph signal reconstruction via Sobolev smoothness}
\newacronym{GRIN}{GRIN}{graph recurrent imputation network}
\newacronym{AWGN}{AWGN}{additive white Gaussian noise}

\begin{document}
\title{Kernel Based Reconstruction for Generalized Graph Signal Processing}

\author{Xingchao Jian~\IEEEmembership{Student~Member,~IEEE}, Wee~Peng~Tay~\IEEEmembership{Senior~Member,~IEEE}, Yonina C. Eldar~\IEEEmembership{Fellow,~IEEE}%
	\thanks{
		This research is supported by the Singapore Ministry of Education Academic Research Fund Tier 2 grant MOE-T2EP20220-0002.  
		
		Xingchao Jian and Wee Peng Tay are with the School of Electrical and Electronic Engineering, Nanyang Technological University, 639798, Singapore (E-mails: \texttt{xingchao001@e.ntu.edu.sg, wptay@ntu.edu.sg}). 
		
		Yonina C. Eldar is with the Faculty of Mathematics and Computer Science,
		The Weizmann Institute of Science, Rehovot 7610001, Israel (E-mail: \texttt{yonina.eldar@weizmann.ac.il}).
	}%
}



\maketitle \thispagestyle{empty}


\begin{abstract}
In generalized graph signal processing (GGSP), the signal associated with each vertex in a graph is an element from a Hilbert space. In this paper, we study GGSP signal reconstruction as a kernel ridge regression (KRR) problem. By devising an appropriate kernel, we show that this problem has a solution that can be evaluated in a distributed way. We interpret the problem and solution using both deterministic and Bayesian perspectives and link them to existing graph signal processing and GGSP frameworks. We then provide an online implementation via random Fourier features. Under the Bayesian framework, we investigate the statistical performance under the asymptotic sampling scheme. Finally, we validate our theory and methods on real-world datasets.
\end{abstract}


\section{Introduction}\label{sect:intro}

\IEEEPARstart{I}{n} real-world signal processing, data is often associated with a network. \Gls{GSP} techniques have been proposed to perform filtering, sampling and reconstruction for this class of signals by accommodating to the network structure \cite{ShuNarFroOrtVan:J13,OrtFroKov:J18}. \gls{GSP} models and exploits the relationship between signals and graphs through the definitions of the \gls{GFT} and frequency. In practice, \gls{GSP} can be utilized to analyze brain signals \cite{MedHuaKar:J18, HuaGolWym:J16}, denoise an image \cite{CheMagTan:J18,YagOzg:J20}, and design recommendation systems \cite{HuaMarRib:J18}. 

Graph signal reconstruction aims to recover the entire graph signal based on observations from a subset of vertices. The major tasks in graph signal reconstruction are designing optimal sampling and recovery strategies \cite{TanEldOrt:J20,TanEld:J20}. When the graph signal is bandlimited, \cite{CheVarSanKov:J15} derived a least squares estimator. 
Based on this estimator, \cite{AniGadOrt:J16} formulates the sampling problem as an optimization problem. Assuming \gls{WSS} and bandlimited signal and \gls{WSS} noise, \cite{ChaRib:J18} studied a greedy sampling scheme, and derived a bound for its recovery \gls{MSE}. The paper \cite{HarTanEld:J23} derived the Wiener filter for graph signal reconstruction under the assumption of \gls{WSS}, while \cite{QiuMaoShe:J17} studied the reconstruction problem for time-varying graph signals. By requiring smoothness in the vertex domain of the graph signals' first-order difference over time, reconstruction is formulated as an optimization problem. This optimization approach is generalized and accelerated in \cite{GirMahGar:J22} through the Sobolev smoothness term. The work \cite{KroRouEld:J22} studied the problem of recovering graph signals from nonlinear measurements.

Kernel-based \gls{GSP} techniques have more flexibility in filtering and reconstruction, since it introduces nonlinearity and generalizes the existing approaches. In \cite{VenChaHan:C20}, the graph signal is modeled as a random nonlinear function of an arbitrary input with a specific covariance structure adapted to the graph, known as a \gls{GPG}. The covariance structure contains a scalar-valued kernel for differentiating the inputs and contains the graph structure for regularizing the smoothness of the random graph signal. The papers \cite{VenChaHan:C18,VenChaHan:J19,EliaGogMar:J22} formulate a learning problem with a graph signal target. Besides the standard \gls{KRR} fitness and regularization terms, this framework imposes smoothness on the output of the training set. The work \cite{EliaGogMar:J21} generalizes the graph-time linear filter \cite[eq.\ (7)]{IsuLeuBan:C16} to a nonlinear predictor via \gls{KRR}. This model assumes the same nonlinear function on every vertex, hence can be made adaptive and distributed by \glspl{RFF}. In the reconstruction problem, \cite{RomMaGia:C16,RomMaGia:J17} design the graph kernel by viewing the graph signal as a function on the vertex set. This approach generalizes the bandlimited graph signal reconstruction method. By implementing the \gls{MKL} strategy, it does not require knowledge of the signal bandwidth.

The aforementioned techniques are developed in terms of the classical \gls{GSP} framework, where each vertex signal is a \emph{scalar}. In practice, the data associated with each vertex can have additional structure. For example, on each vertex, the observation may be a discrete-time signal of length $T$. This scenario is considered in the time-vertex framework \cite{LouFou:C16,GraLouPerRic:J18,LouPer:J19}, where the spatial-time structure is modeled by a Cartesian product graph, and the Fourier transform and filters are then generalized to this graph. To be specific, the Cartesian product graph is constructed by the underlying graph and the cyclic graph with $T$ vertices, the latter of which represents time steps. The data can then be embedded in this product graph as a standard graph signal. This framework is further extended to the \gls{GGSP} framework \cite{JiTay:C18, JiTay:J19, JiaTay:J22}, where each vertex observation is an element from a Hilbert space, which can possibly be infinite-dimensional. An important example is the case where each vertex is associated with a continuous function on a bounded interval. This model allows for analyzing asynchronously sampled signals on each vertex, which is not possible under the time-vertex framework.

In this paper, we explore kernel-based signal reconstruction within the \gls{GGSP} framework. Previous works have developed signal reconstruction methods within the traditional GSP or time-vertex frameworks. However, to the best of our knowledge, no existing reconstruction frameworks exist for the GGSP framework, which considers signals in a \emph{general Hilbert space}. Specifically, when the signal on each vertex is a real-valued function, by utilizing a reasonable kernel, we will be able to reconstruct the signal well as long as the target signal is in the corresponding \gls{RKHS}.

\begin{figure}[!htb]
	\centering
	\includegraphics[width=0.7\linewidth]{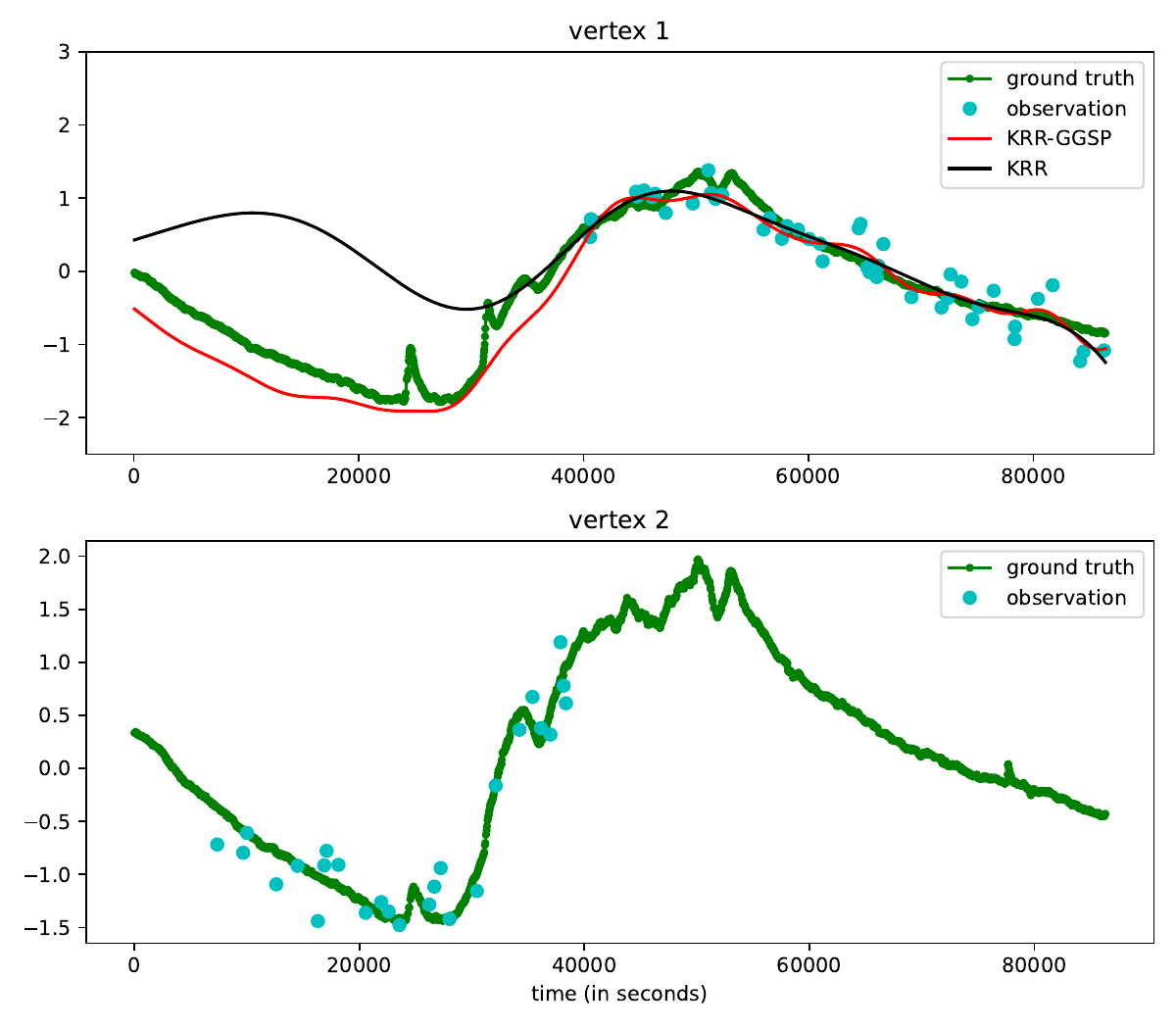}
	\caption{The upper and lower plots represent observations and ground truths from two connected vertices $1$ and $2$, respectively. Green curves are ground truth signals, and the cyan dots represent the observations on each vertex. Note that reconstruction based on the single vertex 1's observations using KRR is much worse compared to the proposed KRR-GGSP approach.}
	\label{fig:illus_intro}
\end{figure}

To motivate our work, consider the Intel lab temperature dataset\footnote{\url{http://db.csail.mit.edu/labdata/labdata.html}} which consists of temperature records from 54 sensors in a lab, collected between February and April of 2004. The ground truth records and incomplete noisy observations on two connected sensors labeled as vertex 1 and 2 are shown in \cref{fig:illus_intro}. Our goal is to reconstruct the signal at vertex 1. In the time interval $[0, 40000]$, there is a lack of observations on vertex 1. As shown in \cref{fig:illus_intro}, the isolated \gls{KRR} method fails to reconstruct this part. On the other hand, our proposed approach, referred to as \gls{KRR}-\gls{GGSP}, utilizes the graph structure to incorporate the observations from a vertex’s neighbor to improve reconstruction. This example motivates the need for a new KRR framework under GGSP, which is the focus of this paper. Unlike the methods under \gls{WSS} or \gls{JWSS} assumptions \cite{HarTanEld:J23,LouPer:J19}, KRR-GGSP does not require knowledge of the \gls{PSD} of the signal, which can be hard to estimate when there are only noisy and incomplete samples in the training set. Further numerical experiments in \cref{sect:numexp} illustrate the utility of the approach presented in this paper.

Our main contributions are the following:
\begin{enumerate}
\item We construct an appropriate kernel and formulate the signal reconstruction in \gls{GGSP} as a \gls{KRR} problem. We interpret it as an extension of existing kernel-based frameworks.
\item We present an online approach for generalized graph signal reconstruction by utilizing \gls{RFF}.
\item We compute the limit and asymptotic upper bound for conditional \gls{MSE} of reconstruction under the Bayesian framework.
\item We present numerical case studies to illustrate the utility of KRR-GGSP in several applications.
\end{enumerate}

This paper is related to our conference paper \cite{JiaTay:C23}, whose goal was to learn a map from a generalized graph signal space to itself in \emph{filtering}. We made use of the tensor product operator-valued kernel to formulate this filtering problem. In this paper, we instead study the \emph{reconstruction} problem for generalized graph signal and our goal is to learn a function from the set of sample points to $\Real$. Here the sample points are pairs of vertices and instances of the vertex function's domain. To achieve this, we consider a real-valued kernel defined on the set of sample points. We make use of the tensor product strategy to form a kernel. 

The rest of this paper is organized as follows. In \cref{sect:prelim}, we formulate the signal reconstruction problem in \gls{GGSP}. In \cref{sect:KRR_GGSP}, we derive the solution to this problem, discuss its interpretation and compare it with existing methods. We also provide an online version of the reconstruction problem. In \cref{sect:stat_analysis}, we analyze the statistical performance of our reconstruction approach under the asymptotic case. In \cref{sect:numexp}, we validate our method on real-world datasets. We conclude in \cref{sect:Conclusion}.

\emph{Notations.} We use plain lower cases (e.g., $x$) to represent scalars and scalar-valued functions. We use bold lower cases (e.g., $\bx$) to represent vectors and vector-valued functions. Note that in this paper, we consider generalized graph signals as scalar-valued functions. Although they are vectors in linear spaces, we use the functional view for ease of explanation. Bold upper cases (e.g., $\bS$) are used to denote operators, including matrices. In particular, we write the $N$-dimensional identity operator or matrix as $\bI_N$. We use calligraphic letters to represent spaces (e.g., $\calX$), except for standard spaces like $\Real$ and $\Nat$, which are the Euclidean space and space of natural numbers, respectively. For a Hilbert space $\calH$, its inner product is $\ip{}{}_{\calH}$ and corresponding norm is $\norm{}_{\calH}$. 
For two random variables (or elements) $x$ and $y$, we write $x\in\sigma(y)$ if $x$ is measurable \gls{wrt} the $\sigma$-algebra generated by $y$. We use $\delta(\cdot,\cdot)$ to denote the Kronecker delta function, which equals $1$ if its two arguments are the same and $0$ otherwise. 
The tensor product is denoted by $\otimes$ and $\diag(\bv)$ is the diagonal matrix with its main diagonal given by the vector $\bv$. The element-wise matrix multiplication is denoted by $\odot$, $(\cdot)\T$ denotes transpose, $(\cdot)^*$ denotes conjugate transpose or the adjoint, and $(\cdot)^\dag$ denotes the pseudo-inverse. We use $[m]$ to represent the set $\set{1,\dots,m}$.


\section{Problem Formulation}\label{sect:prelim}

In this section, we formulate the generalized graph signal reconstruction problem.

Consider a graph $G=(\calV,\calE)$, where $\calV=\set{1,\dots,N}$ is the vertex set, and $\calE\subset\calV\times\calV$ is the edge set. We use $\calN_{d}(v)$ to denote the $d$-hop neighborhood of the vertex $v$ and let $\widebar{\calN}_d(v)= \calN_d(v)\cup\set{v}$. We assume that $G$ is a connected undirected graph with no self-loops. In GSP theory, a typical graph signal is a function mapping from $\calV$ to $\Real$.\footnote{For simplicity, we consider only $\Real$-valued signals instead of $\bbC$-valued signals.} In the \gls{GGSP} framework \cite{JiTay:J19}, the generalized graph signal $f$ is defined as a function from $\calV$ to a separable Hilbert space $\calH$. The generalized graph signal space can then be identified with $\RH$ via the map
\begin{align*}
	f&\mapsto \sum_{n=1}^{N} \be_n\otimes f(n),
\end{align*}
where $\set{\be_n \given n=1,\dots,N}$ is the standard basis of $\Real^N$, i.e., $\be_n$ is the $n$-th column vector of $\bI_N$.

One important case in \gls{GGSP} is where $\calH$ is a function space. Specifically, consider the domain of the functions to be a measure space $(\calT,\calA,\tau)$ and $\calH=L^2(\calT)$. Then, a generalized graph signal $f$ can be identified with the map 
\begin{align*}
	f':\calV\times\calT&\to\Real\\
	(v,\bt)&\mapsto f(v)(\bt).
\end{align*}
Thus, the space of generalized graph signals can be also identified with $L^2(\calV\times\calT)$. In this paper, we will mainly use $L^2(\calV\times\calT)$ to denote the space of generalized graph signals, while references to $\RH$ are used in explanations and proofs. We refer to $\calT$ colloquially as the \emph{time} domain. Readers are referred to \cref{sect:GGSP} and \cite{JiTay:J19} for more details on \gls{GGSP}.

Given noisy observation samples at a subset $\calS \subset \calV\times\calT$ of vertices and time instances, our objective is to recover the generalized graph signal $f$. To avoid cluttered notations, denote $\calJ=\calV\times\calT$. Suppose the sampling set is $\calS = \set{(v_m,\bt_m) \given m=1,\dots,M}\subset\calJ$, and the noisy observations are 
	\begin{align}\label{obs_model}
		y_m = f(v_m,\bt_m) + \epsilon_m,\ m = 1,\dots,M,
	\end{align}
where $\epsilon_m$ are \gls{iid} zero-mean noise with variance $\sigma^2$. In the Bayesian framework, $f$ in \cref{obs_model} is further modeled as a Gaussian process. In this case, we will model $f$ as a random element (cf .\ \cref{subsect:RE_intro}). The noise terms $\epsilon_m$ are assumed to be Gaussian and independent of this process. 

The \gls{GGSP} signal reconstruction problem can be summarized in the following form:
\begin{align}\label{eq:GGSP_general_prob}
	\min_{\tilde{f}\in F(\calJ, \Real)} \sum_{m=1}^M L(\tilde{f}(v_m,\bt_m), y_m) + P(\tilde{f}),
\end{align}
where $F(\calJ, \Real)$ is an appropriate space of functions from $\calJ$ to $\Real$, $L(\cdot)$ is a loss function measuring the fitness of $\tilde{f}$ on the observations. Typical choices include the $\ell_1$ and $\ell_2$ losses. The regularization term $P(\tilde{f})$ imposes a smoothness constraint on $\tilde{f}$ over the vertex and time domains. To design proper $F(\calJ, \Real)$ and $P(\tilde{f})$, we employ the \gls{KRR} technique, which we briefly review in \cref{subsect:KRR_intro}. 

The existing time-vertex methods \cite{QiuMaoShe:J17,GirMahGar:J22} have already addressed the reconstruction problem for time series on graphs. However, these methods are based on the assumption that the signals are evenly sampled with the same sampling rate on all vertices. In contrast, from \cref{eq:GGSP_general_prob}, we observe that our formulation does not require synchronous samples from each vertex and applies even in the case where the sampling frequencies differ across vertices, or where the signal is not evenly sampled. In addition, compared to the time-vertex methods, this formulation is not sensitive to the sampling rate since it makes use of the true time stamps. We refer the reader to the detailed discussion in \cref{subsect:Bayes_interpret}.

\section{KRR Reconstruction in GGSP}\label{sect:KRR_GGSP}

In this section, we derive the KRR reconstruction solution for GGSP. We interpret this method under both deterministic and Bayesian models and connect our method with existing kernel-based frameworks in \gls{GSP} and graph signal reconstruction approaches. We also propose an online approach based on \gls{RFF} that results in a distributed implementation.

To reconstruct a generalized graph signal $f\in L^2(\calJ)$, we use a kernel $k:\calJ\times\calJ\to\Real$ that is the multiplication of two kernels $k_G:\calV\times\calV\to\Real$ and $k_\calT:\calT\times\calT\to\Real$:
\begin{align}\label{eq:mult_ker}
	\begin{aligned}
	k:\calJ\times\calJ&\to\Real\\
	((u,\bs),(v,\bt))&\mapsto k_G(u,v)k_\calT(\bs,\bt).
	\end{aligned}
\end{align}
The \gls{RKHS} associated with the kernel \cref{eq:mult_ker} is $\calH_k = \calH_{k_G}\otimes\calH_{k_\calT}$ \cite[Theorem 13]{BerTho:11}. In this paper, we focus on the case where the matrix $\bK_G:=(k_G(i,j))\in\Real^{N\times N}$ takes the following form (cf.\ \cite[(14)]{RomMaGia:J17}):
\begin{align}\label{eq:Lap_ker}
	\bK_G = \bPhi\diag(r(\lambda_1),\dots,r(\lambda_N))\bPhi\T,
\end{align}
where $\set{\lambda_i}$ are the eigenvalues of the GSO $\bA_G$, $r(\cdot)$ is a non-negative function such that $r(\lambda_1)\geq\dots\geq r(\lambda_N)$,\footnote{Recall that $\set{\lambda_i}$ are indexed in increasing order of graph frequencies. Also note that \cite[(14)]{RomMaGia:J17} uses $r^\dag(\bLambda)$ instead of $r(\bLambda)$ in the definition \cref{eq:Lap_ker}.} and $\bPhi$ is the matrix formed by the eigenvectors of $\bA_G$. When $\calT$ is a subset of Euclidean space, we can usually choose $k_\calT$ as the \gls{RBF} kernel, e.g., $k_\calT(\bs,\bt) = \exp(-\norm{\bs-\bt}_2^2/\gamma)$ (Gaussian kernel) and $k_\calT(\bs,\bt) = \exp(-\norm{\bs-\bt}_1/\gamma)$ (Laplacian kernel), where $\gamma$ is a tunable parameter.

Following the standard \gls{KRR} formulation \cref{eq:KRR_prob}, we specify the reconstruction problem \cref{eq:GGSP_general_prob} as follows:
\begin{align}\label{eq:GGSP_KRR_prob}
 	\hat{f} = \argmin_{\tilde{f}\in\calH_k} \sum_{m=1}^M \abs{\tilde{f}(v_m,\bt_m) - y_m}^2 + \mu \norm{\tilde{f}}_{\calH_k}^2.
\end{align}
Let $\bK(\calS,\calS)=(k((v_m,\bt_m),(v_{m'},\bt_{m'})))_{m,m'=1}^M\in\Real^{M\times M}$ and $\by(\calS)=(y_1,\dots,y_M)\T$. Using the representer theorem, the optimal solution to \cref{eq:GGSP_KRR_prob} is 
\begin{align}\label{eq:GGSP_KRR_sol}
	\begin{aligned}
	\hat{f}&=\sum_{m=1}^Mc_mk(\cdot,(v_m,\bt_m)),\\
	(c_1,\dots,c_M)\T&=(\bK(\calS,\calS)+\mu\bI_M)^{-1}\by(\calS).
	\end{aligned}
\end{align}
Henceforth, we refer to the problem \cref{eq:GGSP_KRR_prob} and its solution \cref{eq:GGSP_KRR_sol} as \gls{KRR}-\gls{GGSP}. By construction \cref{eq:Lap_ker}, $\bK_{G}$ is a polynomial of $\bA_G$ for some degree $L<N$, so that $k_G(u,v)=0$ as long as $u\notin\widebar{\calN}_{L}(v)$. Therefore, the evaluation of $\hat{f}(v,\bt)$ only requires information from $\widebar{\calN}_{L}(v)$:
\begin{align}\label{eq:KRR-GGSP_local}
	\hat{f}(v,\bt)&=\sum_{m=1}^Mc_mk((v,\bt),(v_m,\bt_m))\nn
	&=\sum_{m=1}^Mc_mk_G(v,v_m)k_\calT(\bt,\bt_m)\nn
	&=\sum_{v_m\in\widebar{\calN}_{L}(v)} c_mk_G(v,v_m)k_\calT(\bt,\bt_m).
\end{align}

Note that when $\calT$ is a singleton (i.e., the vertex signal space is one-dimensional), the KRR-GGSP framework degenerates to the \gls{GSP} recovery problem \cite{RomMaGia:J17}. In addition, when $\bK_{G}=\bI_N$, it degenerates to separately solving \gls{KRR} problems on each vertex using the kernel $k_\calT$. To see this, we relabel $\calS$ and $\set{y_m}$ such that $\calS=\bigcup\limits_{v\in\calV}\set{(v,\bt_{i}^{(v)})\given i=1,\dots,M_v}$, $\set{y_m}=\bigcup\limits_{v\in\calV}\set{y_i^{(u)}\given i=1,\dots, M_u}$. We also relabel the coefficients as $c_{i}\tc{v}$, so that
\cref{eq:GGSP_KRR_sol} can be rewritten as
\begin{align*}
	\hat{f}(u,\bt)
	=\sum_{v=1}^N\delta(u,v)\sum_{i=1}^{M_u} c_{i}^{(v)}k_\calT(\bt,\bt_{i}^{(v)})
\end{align*}
for each $u\in\calV$ and $\bt\in\calT$.
Note that $\hat{f}(u,\bt)= \sum\limits_{i=1}^{M_u} c_{i}^{(u)}k_\calT(\bt,\bt_{i}^{(u)})$ and
\begin{align*}
	\norm{\hat{f}}_{\calH_k}^2
	&=\sum_{u=1}^{N}\sum\limits_{i,j=1}^{M_u}c_{i}^{(u)}k_\calT(\bt_{i}^{(u)},\bt_{j}^{(u)})c_{j}^{(u)}
	=\sum_{u=1}^{N}\norm{\hat{f}(u,\cdot)}_{\calH_{k_\calT}}^2.
\end{align*}
Then problem \cref{eq:GGSP_KRR_prob} becomes
\begin{align}\label{eq:sep_KRR}
	\hat{f} = \argmin_{\tilde{f}\in\calH_k} \sum_{u=1}^N \sum_{i=1}^{M_u} \abs{\tilde{f}(u,\bt_i^{(u)}) - y_i^{(u)}}^2 + \mu\sum_{u=1}^{N}\norm{\tilde{f}(u,\cdot)}_{\calH_{k_\calT}}^2,
\end{align}
and each $\hat{f}(u,\cdot)$ can be solved separately using the samples on the vertex $u$.

\subsection{Deterministic Interpretation}\label{subsect:deter_interpret}

In this subsection, we consider the case where $f$ in \cref{obs_model} is deterministic. We make the following assumption.
\begin{Assumption}\label{asp:T_compact}
For the measure space $(\calT,\calA,\tau)$, $\calT$ is a compact metric space, $\calA$ is the Borel $\sigma$-algebra, and $\tau$ is a strictly positive finite Borel measure. The kernel $k_\calT$ is a continuous symmetric positive definite kernel and $\bK_{G}$ is a positive definite matrix.
\end{Assumption}

By Mercer's theorem \cite{SteSco:J12}, there exists an orthonormal sequence $\set{\xi_i\given i\geq 1}$ in $L^2(\calT)$ such that:
\begin{align*}
	\int_\calT k_\calT(\bs,\bt)\xi_i(\bs) \ud \tau(\bs) &= \gamma_i\xi_i(\bt),\\
	\int_\calT \xi_{i}(\bs)\xi_{j}(\bs) \ud \tau(\bs)&= \delta(i,j),\\
	k_\calT(\bs,\bt) &= \sum_{i=1}^\infty \gamma_i \xi_i(\bs)\xi_i(\bt),
\end{align*}
where the sum converges absolutely and uniformly on $\calT$ and $\gamma_i$, $i\geq1$, are non-negative eigenvalues. Since $k_G$ is given by \cref{eq:Lap_ker}, it can be decomposed in the same way:
\begin{align*}
	k_G(u,v) = \sum_{n=1}^N r(\lambda_n)\phi_n(u)\phi_n(v).
\end{align*}
By definition of $k$ in \cref{eq:mult_ker}, we then have
\begin{align*}
	k((u,\bs),(v,\bt)) 
	= \sum_{n=1}^N\sum_{i=1}^\infty r(\lambda_n)\gamma_i\cdot \phi_n(u)\xi_i(\bs)\cdot\phi_n(v)\xi_i(\bt).
\end{align*}
Note that $\set{\phi_n(\cdot)\xi_i(\cdot)\given n=1,\dots,N, i\geq1}$ is a orthonormal sequence in $L^2(\calJ)$. Following the same argument as \cite{Wah:J73}, $\calH_k$ is a subset of $L^2(\calJ)$ where the functions $\tilde{f}$ satisfy the following condition:
\begin{align}\label{eq:H_kcharacterize}
	\begin{aligned}
	& \tilde{f}(v,\bt) = \sum_{n=1}^N\sum_{i=1}^\infty c_{n,i}\cdot \phi_n(v)\xi_i(\bt) \\
	&\ST \norm{\tilde{f}}_{\calH_k}^2 = \sum_{n=1}^N\sum_{i=1}^\infty \frac{c_{n,i}^2}{r(\lambda_n)\gamma_i} <\infty.
	\end{aligned}
\end{align}
By the definition of \gls{JFT} (cf.\ \cref{eq:JFT}), it can be shown that $c_{n,i} = \calF_{n,i}(\tilde{f})$. Therefore, penalizing on $\norm{\tilde{f}}_{\calH_k}$ is the same as penalizing on the energy of $\calF_{n,i}(\tilde{f})$ with weights $\ofrac*{r(\lambda_n)\gamma_i}$. Note that $r(\cdot)$ is non-increasing so that the Fourier coefficients associated with larger graph frequencies are more heavily penalized.

It is worth noting that if we construct $k_\calT$ as 
\begin{align}\label{eq:bandlimit_ker}
	k_\calT(\bs,\bt) = \sum_{i=1}^B \gamma_i\xi_i(\bs)\xi_i(\bt)
\end{align}
for some $B<\infty$, then problem \cref{eq:GGSP_KRR_prob} is equivalent to the bandlimited signal reconstruction in \cite[Section VI.A]{JiTay:J19} with an additional ridge penalty. To see this, we first note that $\calH_k=\spn\set{\phi_n(\cdot)\xi_i(\cdot)\given n=1,\dots,N, i=1,\dots,B}$, i.e., the signal space used for reconstruction is a bandlimited space. Then we substitute \cref{eq:bandlimit_ker} into \cref{eq:H_kcharacterize} to obtain the optimization problem
\begin{align*}
	\hat{f}(v,\bt) = \argmin_{\tilde{f}\in\calH_k} \sum_{m=1}^M \abs{\tilde{f}(v_m,\bt_m) - y_m}^2 + \mu \sum_{n=1}^N\sum_{i=1}^B \frac{c_{n,i}^2}{r(\lambda_n)\gamma_i},
\end{align*}
which coincides with the bandlimited signal reconstruction problem formulated in \cite{JiTay:J19} but with an additional penalty term. This indicates that if $k_\calT$ is not a combination of finite functions, then $\dim(\calH_{k})=\infty$. This implies that the algorithm is able to capture more features than that of bandlimited signals. An example is the Gaussian kernel \cite[Section 4.3.1]{RasWill:05}.

Finally, we discuss the universality of the kernel $k$ in the following theorem.
\begin{Theorem}
	If $k_\calT$ is a universal kernel on $\calT$, then $k$ is universal on $\calJ$.
\end{Theorem}
\begin{proof}
Consider an arbitrary compact set  $\calZ_J\subset\calV\times\calT$, and define $\calZ_v$ such that $\set{v}\times\calZ_v = \calZ_J\cap(\set{v}\times\calT)$. By using the finite-cover definition of a compact set, we note that $\calZ_v$ is compact in $\calT$. Consider an arbitrary $h\in\calC(\calZ_J)$. Let $h_v := h|_{\set{v}\times\calZ_v}$. Due to the universality of $k_\calT$, for any $\epsilon>0$, there exists $h'_v\in \spn\set{k_\calT(\cdot,\bt)\given \bt\in\calZ_v}$ such that $\norm{h'_v-h_v(v,\cdot)}_{\calC(\calZ_v)}<\epsilon$.  Let $h' := \sum_{v=1}^N \delta(v,\cdot)h'_v$. Then, we have $\norm{h'-h}_{\calC(\calZ_J)}<\epsilon$. On the other hand, since $\bK_G$ is positive definite, $\bK_G$ is invertible. Therefore, there exists $\set{a_{v,n}}$ such that $\delta(v,\cdot) = \sum_{n=1}^Na_{v,n}k_G(n,\cdot)$, i.e., $\delta(v,\cdot)\in\spn\set{k_G(n,\cdot)\given n=1,\dots,N}$. By combining the above results, we conclude that $h'\in\calK(\calZ_J)$ and the universality of $k$ follows. 
\end{proof}

\subsection{Bayesian Interpretation}\label{subsect:Bayes_interpret}

We now turn to the Bayesian interpretation where $f\sim\GP(0,k)$ and $\epsilon_m\iid\dist{\calN}[0,\sigma^2]$ in \cref{obs_model}. Let $(\Omega,\calF,\P)$ be the underlying probability space. We regard $\calJ=\calV\times\calT$ as a measure space whose measure is the product measure of counting measure on $\calV$ and the measure $\tau$ on $\calT$. We denote this product measure as $\zeta$. To be specific, $f$ is a stochastic process $\set{f((v,\bt),\omega)\given (v,\bt)\in\calJ,\omega\in\Omega}$. We make the following assumptions:
\begin{Assumption}\label{asp:GRP_assump}\ 
	\begin{enumerate}[i)]
		\item $f((v,\bt),\omega)$ is jointly measurable \gls{wrt} the product measure $\zeta\times\P$.
		\item $f(\cdot,\omega)\in L^2(\calJ)$ for all $\omega\in\Omega$. 
	\end{enumerate}
\end{Assumption}
Under \cref{asp:GRP_assump}, $f$ is a Gaussian random element (cf.\ \cref{thm:Gauss_element}). Henceforth, we abbreviate $f((v,\bt),\omega)$ as $f(v,\bt)$ for simplicity and consistent notations.
First, we note that under the time-vertex framework, the \gls{GP} prior $\GP(0,k)$ is a \gls{JWSS} \gls{GRP}. Consider the case where $\calT=\set{1,\dots,T}$, and $\bK_\calT:=(k_\calT(i,j))\in\Real^{T\times T}$ is a symmetric positive-definite circulant matrix. Then the covariance operator of $\GP(0,k)$ is $\bC_f = \bK_G\otimes\bK_\calT$. Let $\bA_\calH$ be the shift operator 
\begin{align*}
	\bA_\calH(g)(t) = g((t+1)\Mod T),
\end{align*}
which models the case where the vertex observation is a discrete-time signal with $T$ time steps. Since $\bK_\calT$ is a circulant matrix, it commutes with $\bA_\calH$. On the other hand, by the construction of the kernel $k_G$ in \cref{eq:Lap_ker}, we know that $\bK_G$ commutes with $\bA_G$. Therefore, $\bC_f$ commutes with $\bS$, hence $\GP(0,k)$ is a \gls{JWSS} prior. 

\begin{Example}
	The \gls{GP} prior generalizes the \gls{GPG} framework \cite{VenChaHan:C20}, which defined a \gls{GPG} as a vector-valued \gls{GP} whose covariance matrix takes the form
	\begin{align*}
		\cov(\bs,\bt) &= k_\calT(\bs,\bt)\bB(a),\\
		\bB(a) &= (\bI_N + a\bL)^{-2} := (B(a)_{ij}),
	\end{align*}
	where $a>0$ is a parameter. We see that this covariance structure corresponds to a \gls{GP} prior in $L^2(\calJ)$ with $k_G(i,j) = B(a)_{ij}$. The \gls{GPG} also assumes that each observation is $(\bt, \bx)$, where $\bx$ is a complete graph signal, while in \cref{eq:GGSP_KRR_prob} we allow the observed graph signals to be incomplete. Therefore, this generalization allows us to reconstruct the generalized graph signal when the observations come from different subsets of vertices at different instances.
\end{Example}

We next consider the posterior. The observations $\set{(v_m,\bt_m,y_m)}$ are denoted as $\calD_\mathrm{train}$. According to \cref{subsect:KRR_intro}, the \gls{MAP} estimator is given by \cref{eq:GGSP_KRR_sol} with $\mu=\sigma^2$. Since $f$ is a \gls{GP}, \cref{eq:GGSP_KRR_sol} is also the posterior expectation given $\calD_\mathrm{train}$, i.e., $\hat{f}(v,\bt)=\E[f(v,\bt)\given \calD_\mathrm{train}]$. The posterior variance can be calculated by
\begin{align}\label{eq:GP_post_var}
	&\Var(f(v,\bt) \given \calD_\mathrm{train}) \nn
	&= k((v,\bt),(v,\bt))-\bk\T(\bK(\calS,\calS)+\sigma^2\bI_M)^{-1}\bk,
\end{align} 
where $\bk:=(k((v,\bt),(v_1,\bt_1)),\dots,k((v,\bt),(v_m,\bt_m)))\T$. This observation indicates that the time-vertex signal reconstruction approach is a special case of the KRR-GGSP approach.

\begin{Example}\label{exam:GTRSS}
	In the time-vertex signal reconstruction problem, the observed signal $\bX_o\in\Real^{N\times T}$ is an incomplete and noisy observation of the original signal $\bX_r\in\Real^{N\times T}$. The mask matrix is $\bPi_\calS\in\set{0,1}^{N\times T}$. The paper \cite{GirMahGar:J22} formulated the \gls{GTRSS} problem as follows:
	\begin{align}\label{eq:GTRSS}
		\hat{\bX}_r &= \argmin_{\bX\in\Real^{N\times T}}\norm*{\bPi_\calS\odot\bX - \bX_o}_F^2 \nn
		&+ \mu_{\mathrm{TV}} \tr((\bX\bD_h)\T(\bL+\alpha\bI)^{\beta}\bX\bD_h)\nn
		&= \argmin_{\bX\in\Real^{N\times T}}\norm*{\bPi_\calS\odot\bX - \bX_o}_F^2 \nn
		&+ \mu_{\mathrm{TV}} \vect(\bX)\T(\bD_h\bD_h\T)\otimes(\bL+\alpha\bI)^{\beta}\vect(\bX),
	\end{align}
	where $\bD_h$ is the first order difference operator
	\begin{align*}
		\bD_h = 
		\begin{pmatrix}
			-1 & & &\\
			1 & -1 &  & \\
			& 1 & \ddots & \\
			& & \ddots & -1 \\
			& & & 1 
		\end{pmatrix}\in\Real^{T\times (T-1)}.
	\end{align*}
	For ease of further analysis, we slightly modify \cref{eq:GTRSS} to be
	\begin{align}\label{eq:GTRSS_eps}
		\hat{\bX}_r &= \argmin_{\bX\in\Real^{N\times T}}\norm*{\bPi_\calS\odot\bX - \bX_o}_F^2 \nn
		&+ \mu_{\mathrm{TV}} \vect(\bX)\T(\bD_h\bD_h\T + \delta_o\bI)\otimes(\bL+\alpha\bI)^{\beta}\vect(\bX),
	\end{align}
	where $\delta_o>0$. We also assume that  $\diag(\vect(\bPi_\calS))+(\bD_h\bD_h\T)\otimes(\bL+\alpha\bI)^{\beta}$ is full-rank. It can be shown that the solution to \cref{eq:GTRSS_eps} can approximate that of \cref{eq:GTRSS} arbitrarily well as long as $\delta_o$ is small enough. 

	We consider problem \cref{eq:GTRSS_eps} under a Bayesian setting.
	Let the prior of $\vect(\bX_r)$ be a Gaussian random vector with zero mean and covariance $((\bD_h\T\bD_h+ \delta_o\bI)\otimes(\bL+\alpha\bI)^{\beta})^{-1}$. In other words, if we let $k_\calT(s,t)=(\bD_h\bD_h\T+\delta_o\bI)^{-1}_{st}$, and $\bK_{G} = (\bL+\alpha\bI)^{-\beta}$, then $\bX_r = (X_r(v,t))$ is a \gls{GP} with covariance $\cov(X_r(u,s), X_r(v,t)) = k_\calT(s,t)k_G(u, v)$. Suppose the noise is \gls{iid} with variance $\mu_{\mathrm{TV}}$, then the objective function in \cref{eq:GTRSS_eps} is the log-likelihood of the posterior $p(\bX_r|\bX_o)$ (up to a constant):
	\begin{align*}
		\log(p(\bX_r|\bX_o)) &= \log(p(\bX_r,\bX_o)) - \log(p(\bX_o))\\
		 &= \log(p(\bX_o|\bX_r)) + \log(p(\bX_r)) - \log(p(\bX_o))\\
		 &= \ofrac{\mu_{\mathrm{TV}}}\norm*{\bPi_\calS\odot\bX_r - \bX_o}_F^2 \\
		 &+ \vect(\bX_r)\T(\bD_h\bD_h\T+\delta_o\bI)\otimes(\bL+\alpha\bI)^{\beta}\vect(\bX_r) \\
		 &+ \const,
	\end{align*}
  where $\const$ is a constant independent of $\bX_r$. Therefore,
	the solution to this problem is the \gls{MAP} of $\bX_r$ given $\bX_o$. 
	According to the Bayesian interpretation in \cref{subsect:KRR_intro}, this \gls{MAP} estimator $\hat{\bX}_r = (\hat{X}_r(v,t))$ is the solution \cref{eq:GGSP_KRR_sol} of KRR-GGSP where $k_\calT(s,t) = (\bD_h\bD_h\T+\delta_o\bI)^{-1}_{st}$, $\bK_{G} = (\bL+\alpha\bI)^{-\beta}$, and  $\mu=\mu_{\mathrm{TV}}$.
\end{Example}

\begin{figure}
	\centering
	\includegraphics[width=0.3\linewidth, trim=4cm 1cm 4cm 4cm]{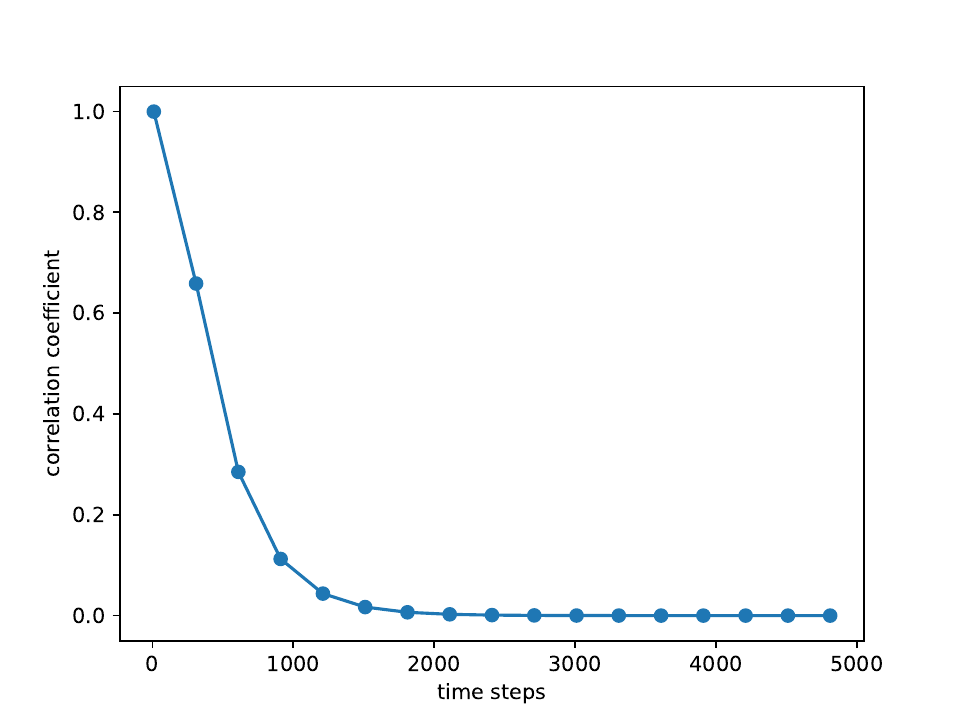}
	\caption{The prior correlation coefficients $\corr(f(0), f(1);T)$ as a function of $T$ with $\delta_o= 10^{-5}$.}
	\label{fig:corr}
\end{figure}

From \cref{exam:GTRSS}, we see that the \gls{GTRSS} problem can be understood as using a specific kernel in the time domain. We note that this kernel depends on the number of discrete time steps, so we denote this kernel as $k_\calT(s,t;T) = (\bD_h\bD_h\T+\delta_o\bI)^{-1}_{st}$, where $s,t\in[T]$. This leads to the problem that the prior distribution assigned to the signal relies on the sampling frequency. For example, consider a signal $f$ on $[a,b]$. Suppose $f$ is evenly sampled with interval length $\ofrac{T-1}$, and we try to recover it using the kernel $k_\calT(s,t;T)$.  According to the Bayesian interpretation (cf. \cref{subsect:KRR_intro}), by using this kernel, we have assumed a prior distribution on $f$. We now examine the cross-correlation of the prior between $f(a)$ and $f(b)$, i.e., $\corr(f(a), f(b);T):=\frac{k_\calT(1,T;T)}{\sqrt{k_\calT(1,1;T)k_\calT(T,T;T)}}$. By calculating this quantity with different values of $T$, we find that it is highly related to the sampling frequency (see \cref{fig:corr}). Specifically, when the sampling frequency is large enough, the prior correlation between $f(a)$ and $f(b)$ tends to zero. Instead, if we use other kernels such as \gls{RBF} kernel, the prior cross-correlation $\frac{k_\calT(a,b)}{\sqrt{k_\calT(a,a)k_\calT(b,b)}}$ does not depend on $T$. This accounts for the failure of \gls{GTRSS} on datasets with high sampling frequency, while \gls{KRR}-\gls{GGSP} with \gls{RBF} kernel works well (see \cref{subsect:Intel_tem_exp}). Therefore, by using more flexible kernels, we can expect better reconstruction results.

\subsection{Online and Distributed Implementation}\label{subsect:online_method}
We now consider the online learning problem where the data stream $\set{(v_m, \bt_m, y_m)}$ arrives sequentially. Upon each arrival of $(v_m, \bt_m)$, the learner is supposed to provide a distributed prediction of $f(v_m, \bt_m)$. After that, $y_m$ is observed and the error is measured by comparing the prediction with $y_m$. The estimator of $f(v_m, \bt_m)$ cannot depend on $y_m$, and the error is used to update the learner for the next prediction. Problem \cref{eq:GGSP_KRR_prob} can be adapted to this setting via \gls{RFF}s when $k_\calT$ is a \gls{RBF} kernel. Denote the columns of $\bK_G^{\ofrac{2}}$ by $[\bp_1,\dots,\bp_N]$, and write $\bp_v=(p_{1,v},\dots,p_{N,v})\T$. For the kernel $k$, the \gls{RFF} can be constructed as 
\begin{align*}
	\bmeta(v,\bt) = \bp_v\otimes\bz(\bt),
\end{align*}
where $\bz(\bt)\in\Real^F$ is the \gls{RFF} of the kernel $k_\calT$, i.e., $\E[\bz(\bs)\T\bz(\bt)] = k_\calT(\bs,\bt)$. By the construction of $\bmeta(v,\bt)$, we have $\E[\bmeta(u,\bs)\T \bmeta(v,\bt)] = k((u,\bs),(v,\bt))$. The reconstructed signal is then $\hat{f}_{\text{RFF}}(v,\bt) = \bc\T\bmeta(v,\bt)$. Problem \cref{eq:GGSP_KRR_prob} is therefore converted to the linear regression problem~\cite[(7)]{LiTonOgl:J21}:
\begin{align}\label{eq:GGSP_RFF}
	\min_{\bc\in\Real^{NF}} q(\bc) = 
	\sum_{m=1}^M (\bc\T\bmeta(v_m,\bt_m)-y_m)^2 + \mu\norm{\bc}_2^2.
\end{align}
Alternatively, if we define $q_m(\bc):=(\bc\T\bmeta(v_m,\bt_m)-y_m)^2 + \frac{\mu}{M}\norm{\bc}_2^2$, then \cref{eq:GGSP_RFF} turns out to be
\begin{align}\label{eq:GGSP_RFF_sep}
	\min_{\bc\in\Real^{NF}} q(\bc)=\sum_{m=1}^{M} q_m(\bc).
\end{align}

The evaluation of $\hat{f}_{\text{RFF}}(v,\bt) = \bc\T\bmeta(v,\bt)$ can be distributed. To illustrate this, write $\bc=(\bc_1\T,\dots,\bc_N\T)\T$ where $\bc_n\in\Real^F$, $n=1,\dots, N$. Since $k_G$ takes the form \cref{eq:Lap_ker}, $\bK_G^{\ofrac{2}}$ can be represented as a polynomial of $\bA_G$ of degree $L_0$, so that $p_{u,v} = 0$ for all $u\notin \calN_{L_0}(v)$. Then for any input $(v,\bt)$, $\bmeta(v,\bt) = (p_{1,v}\bz(\bt)\T, \dots, p_{N,v}\bz(\bt)\T)\T$, $\hat{f}_{\text{RFF}}$ is evaluated by
\begin{align*}
	\hat{f}_{\text{RFF}}(v,\bt) = \sum_{u\in\calN_{L_0}(v)} \bc_u\T p_{u,v}\bz(\bt),
\end{align*}
which only requires information from $\calN_{L_0}(v)$.

Problem \cref{eq:GGSP_RFF} can be solved in an online and distributed way by \gls{SGD}. To be specific, suppose the datastream is $\set{(v_m,\bt_m,y_m)\given m=1,2,\dots}$. At the $m$-th step, we approximate $\nabla q$ with the instantaneous sample $(v_m,\bt_m,y_m)$:
\begin{align*}
	\nabla q_m = 2(\bc\T\bmeta(v_m,\bt_m)-y_m)\bmeta(v_m,\bt_m) + 2\frac{\mu}{M}\bc.
\end{align*}
Note that $y_m-\bc\T\bmeta(v_m,\bt_m) = y_m- \hat{f}_{\text{RFF}}(v_m,\bt_m):=\hat{e}_m$ is the approximation error at the current sample point $(v_m,\bt_m)$. We can update $\bc$ at the $m$-th iteration via 
\begin{align}\label{eq:SGD_update}
	\bc^{(m)} = \bc^{(m-1)}-\theta \nabla q_m
	= \theta_1\bc^{(m-1)} + \theta_2 \hat{e}_m \bmeta(v_m,\bt_m),
\end{align}
where $\theta,\theta_1,\theta_2>0$. According to \cite[Theorem 6.11]{GarGow:J23}, the convergence rate of \gls{SGD} is linear when $\mu>0$. Since $\bp_{v_m}$ only has non-zero entries in $\calN_{L_0}(v_m)$, and $\hat{e}_m$ can be evaluated in a distributed way, we see that \cref{eq:SGD_update} is an online and distributed update. This is always achievable when $k_\calT$ is a \gls{RBF} kernel.

\section{Conditional MSE of KRR-GGSP in the Bayesian framework}\label{sect:stat_analysis}

In this section, we consider $f\sim\GP(0,k)$, i.e., the Bayesian framework considered in \cref{subsect:Bayes_interpret}. We derive the \gls{MSE} of the estimate given by KRR-GGSP at a particular node $v_0 \in \calV$ and time $\bt_0 \in \calT$, conditioned on an observation set $\set{(v_m,\bt_m,y_m)\given m=1,\dots,M}$. To be specific, we analyze 
\begin{align}
	\begin{aligned}\label{eq:post_var_point}
	&\Var(f(v_0,\bt_0)\given \set{(v_m,\bt_m,y_m)}) \\
	&= \E[(\hat{f}(v_0,\bt_0) - f(v_0,\bt_0))^2\given \set{(v_m,\bt_m,y_m)}]
	\end{aligned}
\end{align}
under the scenario when the noise energy is unknown, and the MSE is hard to compute when $M\to\infty$ as it involves taking the inverse of the kernel matrix of the observations. We study the dependence of the MSE on the graph structure when a subset of vertices have dense observation samples ($M\to\infty$). The asymptotic MSE and its upper bound can be used as a criterion to choose an optimal sampling vertex set.

We consider the case where an infinite number of samples are observed to infer $f(v_0,\bt_0)$. Note that if we allow uniform sampling on every vertex with an ever-growing sample size, then it is known that the posterior variance will uniformly converge to $0$ \cite{KoePfa:J21}. In order to examine the effect of leveraging information from other vertices in \gls{KRR}-\gls{GGSP}, we consider the case where there are no available sample points on $\set{v_0}\times \calT$, and the value of $f(v_0,\bt)$ is to be estimated. 

\begin{figure}[!htb]
	\centering
	\includegraphics[width=0.4\linewidth, trim=7cm 2cm 10cm 2cm, clip]{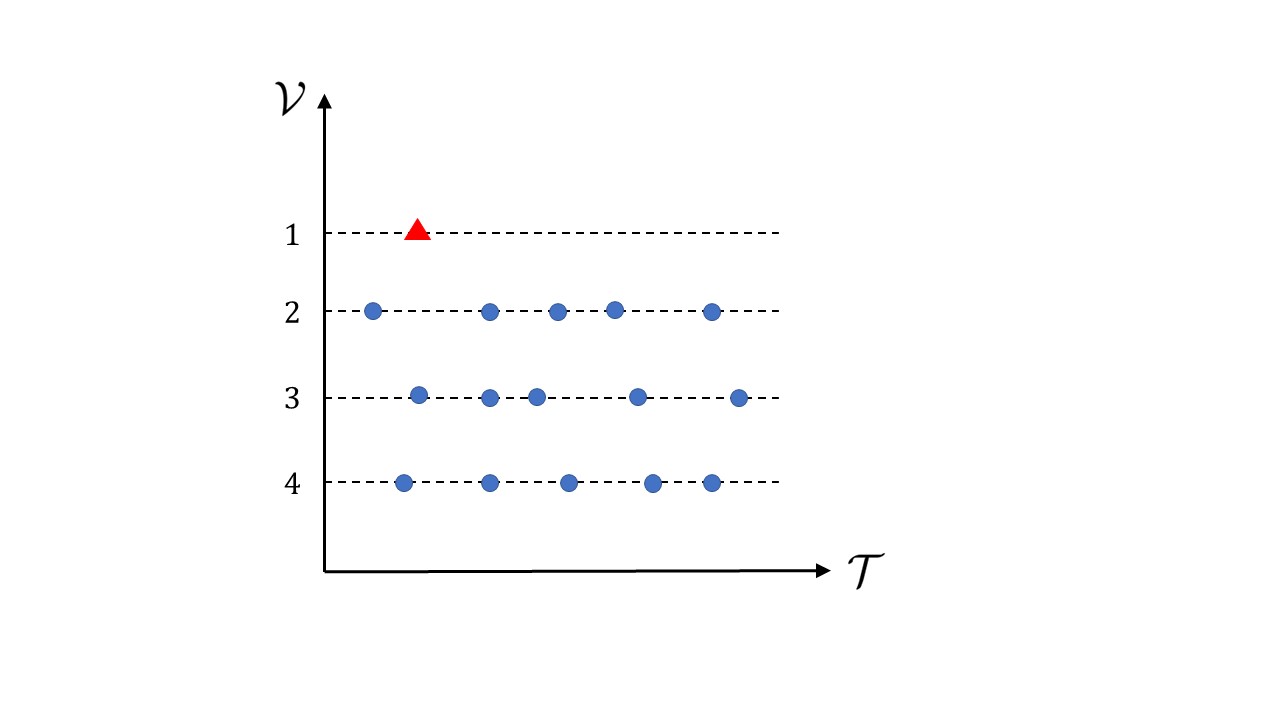}
	\caption{The uniform exclusive sampling scheme with $M_0=5$. The blue circles denote $\calS(M_0)$, and the red triangle is $(v_0,\bt_0)$.}
	\label{fig:uniform-sample}
\end{figure}

Mathematically, let $\calS(v;M_0)$ be a set of $M_0$ samples \gls{iid} from $\dist{Unif}[\set{v}\times\calT]$, where $v\in\set{v_0}\setcomp$. The sample set $\calS(M_0)$ is then obtained by $\calS(M_0)=\bigcup\limits_{v\in\set{v_0}\setcomp}\calS(v;M_0)$. This sampling scheme is illustrated in \cref{fig:uniform-sample}, and we call it \emph{uniform exclusive sampling}. In practice, this scheme mimics the scene where only limited knowledge can be obtained from a certain vertex, and an inference for that is desired.

For ease of notation, we define $\calJ_S:= \set{v_0}\setcomp\times\calT$. We write $\by(M_0)$ to represent the observations $\by(\calS(M_0))$ from the sampling set $\calS(M_0)$, and $\bz$ to represent the restriction of $f$ on $\calJ_S$. We analyze $\Var(f(v_0,\bt_0)\given \by(M_0))$ from two aspects: first, in \cref{thm:limit_var} we analyze the integration of $\Var(f(v_0,\bt)\given \by(M_0))$ over $\bt$; then in \cref{thm:var_bound} we provide an asymptotic upper bound for $\Var(f(v_0,\bt_0)\given \by(M_0))$.

Let $\calT_0$ be a subset of $\calT$. We consider the following integration
\begin{align}\label{eq:int_var}
	\int_{\calT_0} \Var(f(v_0,\bt)\given \by(M_0))\ud \tau(\bt),
\end{align}
which represents the conditional \gls{MSE} of the \gls{KRR}-\gls{GGSP} estimator over $\calT_0$. Let $\bx_0$ be the restriction of $f$ on $\set{v_0}\times\calT_0$, and $\bC_{\bx_0 |\by}:=\Cov(\bx_0\given \by(M_0))$. Note that \cref{eq:int_var} can be equally written as $\tr(\bC_{\bx_0 |\by})$ (cf .\ \cref{eq:tr_is_int}). Based on this observation, we analyze the asymptotic behavior of $\bC_{\bx_0 |\by}$.

We compute the covariance operators  $\bC_{\bz\bz}$  and $\bC_{\bz\bx_0}$ for later use:

\begin{align}\label{eq:cov_operators}
	\begin{aligned}
	\bC_{\bz\bz}:L^2(\calJ_S)&\to L^2(\calJ_S)\\
	g(\cdot)&\mapsto\int_{\calJ_S} k(\cdot,\bxi)g(\bxi) \ud \zeta(\bxi)\\
	&= \int_{\calT} \sum_{u\in\set{v_0}\setcomp} k_G(v,u)k_\calT(\bt,\bs)g(u,\bs)\ud \tau(\bs),\\
	\bC_{\bz \bx_0}:L^2(\calT_0)&\to L^2(\calJ_S) \\
	g(\cdot)&\mapsto \int_{\calT_0}k_G(v_0,v)k_\calT(\bt,\bs)g(v_0,\bs) \ud \tau(\bs).
	\end{aligned}
\end{align}

Define the integral operators
\begin{align*}
	\bH: L^2(\calT)&\to L^2(\calT)\\
	g(\cdot)&\mapsto\int_{\calT}k_\calT(\bt,\bs)g(\bs)\ud \tau(\bs),\\
	\bH_0: L^2(\calT_0)&\to L^2(\calT)\\
	g(\cdot)&\mapsto\int_{\calT_0}k_\calT(\bt,\bs)g(\bs)\ud \tau(\bs).
\end{align*}
Then we have 
\begin{align}\label{eq:cov_operator_forms}
	\begin{aligned}
		\bC_{\bz\bz} = \bK_{G,**} \otimes \bH,\\
		\bC_{\bz\bx_0} = \bk_{G,0*}\otimes\bH_0.
	\end{aligned}
\end{align}

Intuitively, when $M_0$ tends to infinity, the situation can be interpreted as $f$ on $\calJ_S$ is known and can be utilized for inference. We formally address this in the following theorem:
\begin{Theorem}\label{thm:limit_var}
	Under \cref{asp:T_compact}, the limit posterior covariance of $f(v_0,\bt)$ over $\calT_0$  satisfies
	\begin{align*}
		\lim_{M_0\rightarrow\infty} \bC_{\bx_0|\by} = \bC_{\bx_0|\bz}
	\end{align*}
in trace norm. In other words, the conditional variance 
\begin{align*}
	\tr(\bC_{\bx_0|\by}) &= \E[\norm{\bx_0-\E[\bx_0\given\by(M_0)]}^2\given\by(M_0)] \\
	&= \int_{\calT_0}\Var(f(v_0,\bt) \given \by(M_0)) \ud \tau(\bt)
\end{align*}
converges:
	\begin{align}\label{eq:limit_var}
		\begin{aligned}
		\lim\limits_{M_0\rightarrow\infty}\tr(\bC_{\bx_0|\by})
		&=\tr(\bC_{\bx_0|\bz})\\
		&=\tr(\bC_{\bx_0}) - \tr(\bC_{\bx_0\bz}\bC_{\bz\bz}^\dag\bC_{\bx_0\bz}^*)\\
		&= \int_{\calT_0} \Var(f(v_0,\bt) \given \bz)\ud \tau(\bt).
		\end{aligned}
	\end{align}
\end{Theorem}
\begin{proof}
	See \cref{sect:proof:thm:limit_var}.
\end{proof}

From \cref{thm:limit_var} we know the limiting posterior variance given an infinite number of sample points. This result can also be applied when only a subset of vertices have dense samples. In that case, the \gls{RHS} of \cref{eq:limit_var} becomes an asymptotic upper bound by letting $\bz$ be the restriction of $f$ on the vertices with dense samples. Moreover, we can get a rough idea of the behavior of $\Var(f(v_0,\bt_0) \given \by(M_0))$ if we consider the following sequence of continuous functions
\begin{align*}
	\rho_{M_0}(\alpha) := 
	\begin{cases}
		\ofrac{\alpha}\int_{B(\bt_0,\alpha)} \Var(f(v_0,\bt) \given \by(M_0)) \ud \tau(\bt),&\alpha>0\\
		\Var(f(v_0,\bt_0) \given \by(M_0)),&\alpha=0
	\end{cases}
\end{align*}
where $B(\bt_0,\alpha)$ is the open ball centered at $\bt_0$ with measure $\alpha$. Specifically, by \cite[Theorem 3]{KoePfa:J21} we note that $\rho_{M_0}(\alpha)$ is a monotonic sequence, i.e., $\rho_{M_0}(\alpha)\leq\rho_{M_0'}(\alpha)$ if $M_0>M_0'$. According to \cref{thm:limit_var}, the limit function of $\rho_{M_0}(\alpha)$ is 
\begin{align*}
	\rho(\alpha) &= \lim_{M_0\rightarrow\infty} \rho_{M_0}(\alpha) = \ofrac{\alpha} \int_{B(\bt_0,\alpha)}\Var(f(v_0,\bt) \given \bz)\ud \tau(\bt)
\end{align*}
when $\alpha>0$, and
\begin{align*}
	\rho(0) &= \lim_{M_0\rightarrow\infty} \Var(f(v_0,\bt_0) \given \by(M_0)).
\end{align*}
Therefore, if we assume that the limit function of $\rho_{M_0}(\alpha)$ is continuous \gls{wrt} $\alpha$ and $\Var(f(v_0,\bt) \given \bz)$ is continuous \gls{wrt} $\bt$, then $\rho(0) = \lim\limits_{\alpha\rightarrow0} \rho(\alpha) = \Var(f(v_0,\bt_0) \given \bz)$, i.e., 
\begin{align}\label{eq:var_point_lim_cond}
	\lim_{M_0\rightarrow\infty} \Var(f(v_0,\bt_0) \given \by(M_0)) = \Var(f(v_0,\bt_0) \given \bz).
\end{align}

From \cref{eq:var_point_lim_cond} we know that, although $\Var(f(v_0,\bt_0) \given \by(M_0))$ is random due to the randomness of $\calS(M_0)$, its limit $\Var(f(v_0,\bt_0) \given \bz)$ is a deterministic quantity when $M_0\to\infty$. In addition, it can be shown by \cref{lem:EVVE} that 
\begin{align*}
	&\Var(f(v_0,\bt_0) \given f(\calQ)) \\
	&= \Var(f(v_0,\bt_0) \given \bz) + \Var(\E[f(v_0,\bt_0) \given \bz]\given f(\calQ))\\
	&\geq \Var(f(v_0,\bt_0) \given \bz),
\end{align*}
for arbitrary finite set $\calQ\subset\calJ_S$. Therefore, according to \cref{eq:var_point_lim_cond}, $\Var(f(v_0,\bt_0) \given f(\calQ))$ can always serve as an upper bound for $\Var(f(v_0,\bt_0) \given \by(M_0))$ when $M_0$ is large enough. Since $\calQ$ is finite, $\Var(f(v_0,\bt_0) \given f(\calQ))$ may be numerically computed. In contrast, We note that the quantities in \cref{eq:limit_var} involve the pseudo-inverse of a possibly infinite-rank operator, which may be difficult to numerically compute. Consider the case when $\calQ=\calN_{d}(v_0)\times\set{\bt_0}$. Let $N_d := \abs{\calN_d(v_0)}$. For simplicity, we introduce the following notations:
\begin{align*}
	\bk_G(v_0,\calN_d) &:= (k_G(v_0,v))_{v\in\calV\backslash\set{v_0}}\in\Real^{N_d}\\
	\bK_G(\calN_d,\calN_d)&:=(k_G(u,v))_{u,v\in\calV\backslash\set{v_0}}\in\Real^{N_d\times N_d}\\
	l(v_0,d)&:=k_G(v_0,v_0)\\
	&-\bk_G(v_0,\calN_d)\T\bK_G(\calN_d,\calN_d)^{-1}\bk_G(v_0,\calN_d),
\end{align*}
so that
\begin{align*}
	\Var(f(v_0,\bt_0) \given f(\calQ)) = k_\calT(\bt_0,\bt_0)l(v_0,d).
\end{align*}
To provide an explicit upper bound for \cref{eq:post_var_point}, we derive an asymptotic bound with a convergence rate for the posterior variance which is locally computable.
\begin{Theorem}\label{thm:var_bound}
	Suppose $\calT$ is a compact subset of $\Real^D$ whose boundary set has measure zero, and $\bt_0$ is an interior point of $\calT$. Suppose $k_\calT$ is Lipschitz continuous on $\calT$. Let $c_0$ be an arbitrary number in $(0,1)$, $d\in\bbN_+$, then we have
	\begin{align*}
		\Var(f(v_0,\bt_0) \given \by(M_0))&\leq k_\calT(\bt_0,\bt_0)l(v_0,d)\\
		&+(C_1c_0^{-1}+ C_{2}c_0^2)M_0^{-\ofrac{3D+1}} \\
		&+ C_3c_0M_0^{-\frac{2}{3D+1}}
	\end{align*}
	with probability at least
	\begin{align*}
		\parens*{1-\ofrac{2}\frac{1}{(1-c_0)^2C_DM_0^{\ofrac{3D+1}}}}^{N_d}.
	\end{align*}
\end{Theorem}
\begin{proof}
	The proof of \cref{thm:var_bound} is included in \cref{proof:thm:limit_var} in the supplementary.
\end{proof}
We note that when $k_\calT$ is \gls{RBF} kernel, $k_\calT(\bt_0,\bt_0)l(v_0,d)$ only depends on the graph structure. In other words, if we are allowed to select a subset of vertices $\calV'\subset\calV$ to recover the signal on $v_0$, then it is preferred that the subgraph with vertex set $\calV'\bigcup\set{v_0}$ has a small $l(v_0,d)$.

\section{Numerical Experiments}\label{sect:numexp}

In this section, we conduct experiments to illustrate the theory and methods of the \gls{KRR}-\gls{GGSP} approach. In the experiments, $\calT$ is an interval, and the target signal is a function on $\calV\times\calT$. In the datasets, the target signal is downsampled on every vertex. We aim to reconstruct the target signal from the randomly selected samples with additive noise. We compare the following algorithms in the experiments:
\begin{enumerate}
	\item\label[method]{mt:krr-ggsp} \gls{KRR}-\gls{GGSP}. We reconstruct the signal using \cref{eq:GGSP_KRR_prob} with the tensor product kernel \cref{eq:mult_ker}. We set $\bK_{G}= a(\bL - \lambda_N\bI)^2 + b\bI$ such that
	\begin{align}\label{eq:quad_graph_ker_exp}
			a(\lambda_1-\lambda_N)^2 +b = 1,
	\end{align}
	and $0\leq b\leq1$ is a tunable parameter. This parameter setting ensures that $1 = r(\lambda_1)\geq\dots\geq r(\lambda_N) = b$ (cf .\ \cref{eq:Lap_ker}). We set $k_\calT$ to be the \gls{RBF} kernel $k_\calT(s,t) = \exp(-{\abs{s-t}^2}/{\gamma})$, where $\gamma$ is a tunable parameter.
	
	\item Isolated \gls{KRR}. We recover the signal on each vertex separately using \gls{KRR} (cf .\ \cref{eq:KRR_prob} and \cref{eq:sep_KRR}). In \cref{sect:KRR_GGSP}, we have shown that this method is equivalent to using $\bK_G=\bI$ in \gls{KRR}-\gls{GGSP}, i.e., fixing $b=1$ in \cref{eq:quad_graph_ker_exp}. 
	\item \gls{GTRSS}. We recover the signal using \cref{eq:GTRSS}, where $\mu_{\mathrm{TV}}$, $\alpha$ and $\beta$ are tunable parameters. 
	\item \Gls{GRIN}. We implement this method using the Spatiotemporal library 
	\cite{Cini_Torch_Spatiotemporal_2022}. 
\end{enumerate}

\subsection{ECoG Dataset}\label{subsect:ecog_exp}

We test the reconstruction performance of \gls{KRR}-\gls{GGSP} on an ECoG multivariate time series dataset.\footnote{\url{https://math.bu.edu/people/kolaczyk/datasets.html}} 
This dataset contains measurements from 76 electrodes on an epilepsy patient during both ictal and pre-ictal periods \cite{KraKolKir:J08}. We make use of the data from 2 ictal periods. Each period lasts 10 seconds with a sampling rate of $400$ Hz. Therefore, the dataset we use is a $76\times8000$ matrix. We use the last $320$ time steps for testing and the $160$ time steps before the test set for training. We add \gls{AWGN} to the dataset and randomly mask the data so that both training and test sets are incomplete and noisy. We set the noise energy of \gls{AWGN} to be $1\%$ of the signal energy. We test the recovery performances of \gls{KRR}-\gls{GGSP}, \gls{GTRSS}, isolated \gls{KRR} and \gls{GRIN} on this dataset.

Except for the isolated \gls{KRR} method, all other methods rely on a graph structure. To construct the graph, we first use the isolated \gls{KRR} to roughly reconstruct the unknown signal values on 160 time steps in the training set, and then calculate the correlation coefficients of these recovered data. We regard two electrodes as connected if the correlation coefficients between them are larger than $0.5$. We set the edge weights to be the correlation coefficients. For \gls{GRIN}, the training set is used for model training and validation. Besides the small training set with $160$ time steps, we also show its performance trained on all available training data from the dataset, i.e., 7680 time steps. For other methods, the training set is used for tuning parameters. The recovery performance is measured by the relative error 
\begin{align}\label{eq:RMSE}
	\frac{\E[(f(v,t)-\hat{f}(v,t))^2]}{\E[f(v,t)^2]}.
\end{align}

\begin{figure}[!htb]
	\centering
	\includegraphics[width=0.6\linewidth, trim=0.5cm 0.3cm 0.5cm 1.4cm, clip]{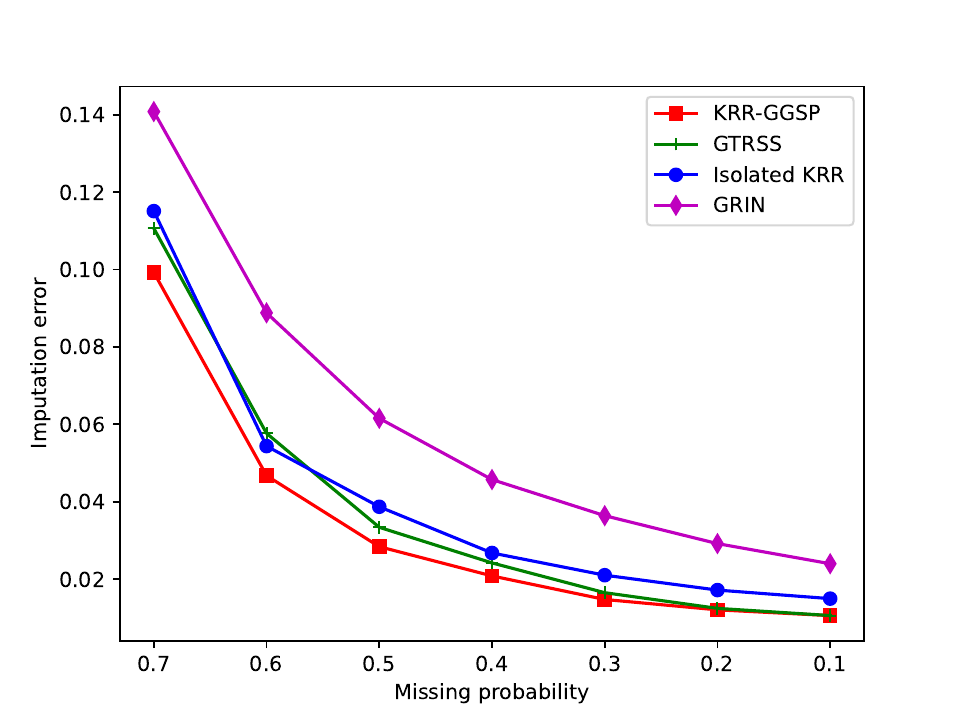}
	\caption{Reconstruction performances on ECoG dataset. Each point in the figure is obtained by 20 repetitions.}
	\label{fig:epilep}
\end{figure}

The recovery results are shown in \cref{fig:epilep}. 
We observe that \gls{KRR}-\gls{GGSP} shows good recovery results and outperforms other methods. Since \gls{KRR}-\gls{GGSP} has a tunable kernel in the time domain, it shows better performance than \gls{GTRSS}. This effect can be better observed in \cref{subsect:Intel_tem_exp}. The isolated \gls{KRR} method has a tunable kernel, but it is not able to take advantage of the graph structure, hence is outperformed by \gls{KRR}-\gls{GGSP}. Here, we show the performance of \gls{GRIN} trained with $7680$ time steps. We remark that the deep learning method \gls{GRIN} requires a sufficiently large training set to obtain reasonable results. When the training set is as small as $160$ time steps, \gls{GRIN} does not yield reasonable reconstruction results.

\subsection{Intel-lab Temperature Data}\label{subsect:Intel_tem_exp}

We test the reconstruction performance of \gls{KRR}-\gls{GGSP} on the Intel lab temperature dataset illustrated in \cref{fig:illus_intro}. In this experiment, we use the data from the first and second days. Since there are $86400$ seconds in a day, the entire dataset we use is a $54\times 172800$ matrix. Here we remark that since the sampling rate of each sensor is much smaller than $1$ Hz and not uniform, only $1.93\%$ of the entries are non-null. Therefore, this dataset is very sparse. We identify the temperature records outside the upper $99.92\%$ quantile and lower $0.001\%$ quantile as outliers and discard them. We subtract the mean value of all observed temperature records from the dataset. We treat each sensor as a vertex and construct a 5-NN graph using their locations. We use half of the first day's records for training and the second day's for testing. As in \cref{subsect:ecog_exp}, we add \gls{AWGN} to the data and assign a random mask. In this experiment, the noise energy is set to be $5\%$ of the signal energy. We compare the methods as described in \cref{subsect:ecog_exp} with performance measurement \cref{eq:RMSE}.

\begin{figure}[!htb]
	\centering
	\includegraphics[width=0.6\linewidth, trim=0.5cm 0.3cm 0.5cm 1.4cm, clip]{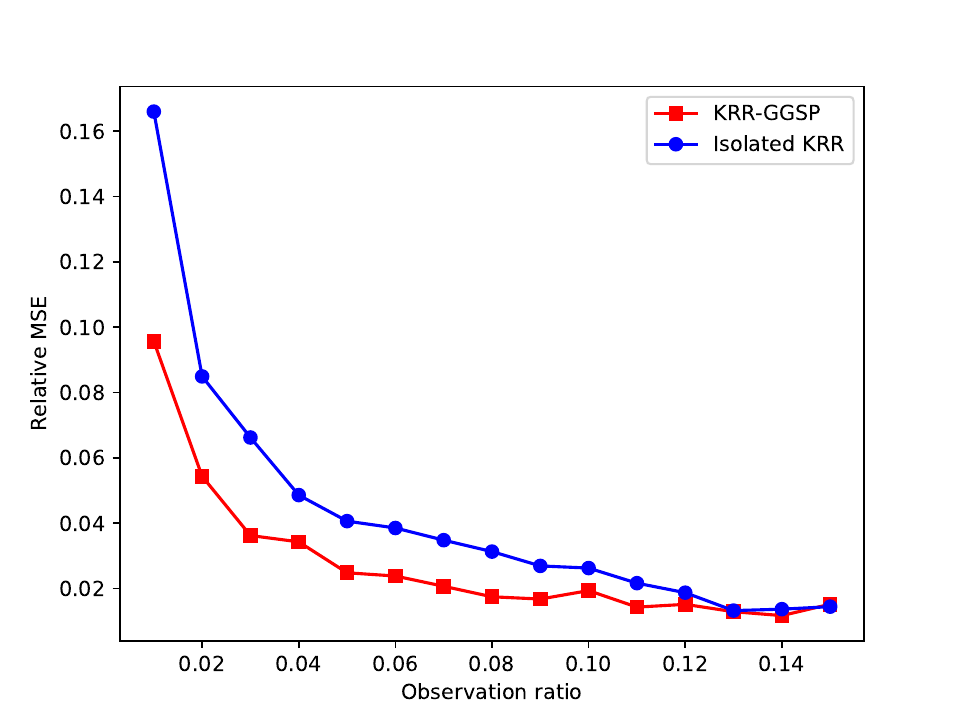}
	\caption{Reconstruction error under different proportions of samples to be used for reconstruction. Each point in the figure is obtained by 10 repetitions.}
	\label{fig:temperature_reconstruct}
\end{figure}

From the result in \cref{fig:temperature_reconstruct}, we observe that \gls{KRR}-\gls{GGSP} outperforms the isolated \gls{KRR}. On this dataset, \gls{GRIN} and \gls{GTRSS} fail to yield reasonable results. For example, when the observation ratio is $0.15$, \gls{GTRSS} has relative \gls{MSE} around $0.8$, and \gls{GRIN} has relative \gls{MSE} around $1.0$. For \gls{GRIN}, this is mainly due to the sparsity of the available data in the dataset. For \gls{GTRSS}, this is due to the improper prior assumption on the dataset.

\subsection{COVID-19 Case Prediction}

We use the online reconstruction method in \cref{subsect:online_method} to predict COVID-19 cases using only historical data. We use the data from The New York Times, based on reports from state and local health agencies\footnote{\url{https://github.com/TorchSpatiotemporal/tsl}}. From this dataset, we retrieve the records from California's 58 counties, starting from the first day when all counties have cases reported so that there are 886 days in total. We treat each county as a vertex and connect them if they are adjacent geographically. We set the datastream and prediction rule as follows: on each date $t$, we randomly choose a subset of vertices $\calV_S=\set{v_1,\dots,v_Q}\subset\calV$ such that the learner is assumed to have access to $\by(\calV_S\times\set{t})$. Besides, for each date $t$, the sample points $\set{(v_i, t, y(v_i,t))}$ are observed sequentially, one datum at a time. 

We compare the online \gls{KRR}-\gls{GGSP} with several existing online and distributed reconstruction methods. The implementation details are the following:
\begin{enumerate}
	\item Online \gls{KRR}-\gls{GGSP}. For each $(v_i,t)\in\calV_S\times\set{t}$, we first calculate the prediction $\hat{f}_\text{RFF}(v_i,t)$. Then we compute the error $\hat{e}_i=y(v_i,t)-\hat{f}_\text{RFF}(v_i,t)$, and update the predictor by \cref{eq:SGD_update}. Then for each $(v_j,t)\in\calV_S\setcomp\times\set{t}$, we also make predictions and compute the error, but will not update the predictor since the learner is not supposed to have access to the observations on them. We set $\bK_{G}=g(\bL)^2$, where $g$ is a polynomial of degree one such that $g(\lambda_1)=1, g(\lambda_N)=0.4$. We let $k_\calT(s,t)=\exp(-{(s-t)^2}/{\gamma})$, where $\gamma$ is an adjustable parameter. We set the dimension of $\bz(t)$ to be $60$.  
	\item Online isolated \gls{KRR}. This is implemented by letting $\bK_{G}=\bI$ in the online \gls{KRR}-\gls{GGSP} method.
	\item Online \gls{GTRSS}. This method is a generalization of \cite[(35)]{QiuMaoShe:J17}, by replacing $\bL$ with $(\bL+\alpha\bI)^\beta$. Let $\hat{\boldf}_{t}^{l}\in\Real^{N}$ be the estimation of ${\boldf}_{t} = (y(1,t),\dots,y(N,t))\T$ after observing $l$ samples on date $t$. The samples are denoted by $\by_t^{l}\in\Real^{N}$ such that the unobserved entries are zero. We write $\boldm_{t}^l$ to denote the mask after observing $l$ samples on date $t$. Let $\hat{\boldf}_{t-1}$ be the estimation of ${\boldf}_{t-1}$ after observing all available samples on date $t-1$. Then the update rule goes as follows:
	\begin{align}
		\hat{\boldf}_{t}^{l} &= \hat{\boldf}_{t}^{l-1} - \mu(\boldm_t^{l}\odot\hat{\boldf}_{t}^{l-1} - \by_t^{l}) \nn
		&- \mu\lambda(\bL+\alpha\bI)^\beta(\hat{\boldf}_{t}^{l-1} - \hat{\boldf}_{t-1}).
	\end{align}
	When the $l+1$-th sample arrives, we evaluate the error $\hat{e}_{l+1} = y(v_{l+1},t) - \hat{f}(v_{l+1},t)$, where $\hat{f}(v_{l+1},t)$ is the $v_{l+1}$-th entry of $\hat{\boldf}_{t}^{l}$. $\lambda, \mu, \alpha$ and $\beta$ are adjustable parameters in this method.
\end{enumerate}

\begin{figure}[!htb]
	\centering
	\includegraphics[width=0.6\linewidth, trim=0.5cm 0.3cm 0.5cm 1.4cm, clip]{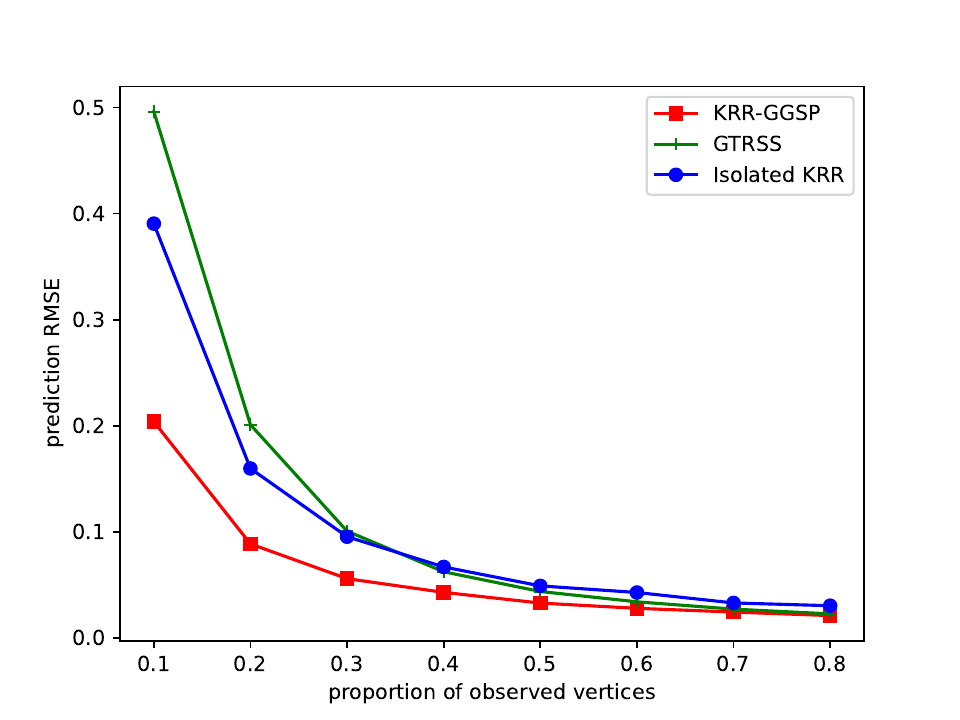}
	\caption{Prediction error under different proportions of vertices to be sampled for learning. Each point in the figure is obtained by 10 repetitions.} 
	\label{fig:COVID_predict}
\end{figure}

We show the best performance of the methods with different parameters in \cref{fig:COVID_predict}. The error measurement is \cref{eq:RMSE}. We observe that the online \gls{KRR}-\gls{GGSP} method outperforms other online and distributed methods. We also tested the ARMA method on each vertex, but due to the missing values, it usually fails to converge and yields unstable results. For example, when the proportion of observed vertices is $80\%$, the ARMA$(2,0,2)$ model fails to converge on about $29\%$ vertices, and the prediction error on each vertex varies from $0.004$ to $665\times 10^4$.

\section{Conclusion}\label{sect:Conclusion}

In this paper, we devised a signal reconstruction approach for \gls{GGSP}, yielding a predictor that can be computed in a distributed fashion. We interpreted this approach in both deterministic and Bayesian aspects and cast it as an extension of existing frameworks. In the former case where the signal is a deterministic function, we showed that the approach imposes smoothness on the reconstructed signal. In the latter case, the signal is regarded as a \gls{GP}, and we analyzed its moments. By utilizing \gls{RFF}, the reconstruction approach can be implemented online, and the evaluation is still distributed.

We provided statistical analysis on the predictor. Under the uniform exclusive sampling scheme, we derived the limit of the posterior variance and provided a numerically computable upper bound for it. We verified the KRR-GGSP approach by numerical experiments. By testing KRR-GGSP against existing methods on real datasets, we validated that introducing the graph structure and the product kernel improves reconstruction performance.

\appendices

\section{Preliminaries: GGSP}\label[Appendix]{sect:GGSP}

In \gls{GSP} theory, typical choices of the \gls{GSO} are the adjacency matrix, Laplacian matrix $\bL$, and their normalized versions. We assume a normal \gls{GSO} denoted as $\bA_G$. Let $\bA_G = \bPhi\bLambda\bPhi\T$ be the eigendecomposition of $\bA_G$, where $\bPhi = [\bphi_1,\dots,\bphi_N]$ consists of orthonormal eigenvectors and $\bLambda=\diag(\lambda_1,\dots,\lambda_N)$. Without loss of generality, we assume that $\set{\lambda_i}$ is indexed in increasing order of the graph frequencies, i.e., $\bphi_N$ is the eigenvector with the highest frequency. The \gls{GFT} is then defined as the Euclidean inner product with the orthonormal basis $\bPhi$, i.e., the operator $\bPhi\T$. In \gls{GGSP}, due to the additional structure in $\calH$, we further assume a shift operator (compact linear transformation) $\bA_\calH$ on $\calH$. The shift operator $\bS$ on $\RH$ is then defined as $\bS := \bA_G\otimes\bA_\calH$. In $L^2(\calJ)$, $\bS$ operates as follows: 
\begin{align*}
	\bS:L^2(\calJ)&\to L^2(\calJ)\\
	f(v,\bt)&\mapsto \bS(f)(v,\bt)=\sum_{n=1}^{N} \bA_G(v,n)\bA_\calH(f(n,\cdot))(\bt),
\end{align*}

Suppose we are given a complete orthonormal basis $\set{\psi_i \given i\geq1}\subset L^2(\calT)$. On the space $L^2(\calJ)$, the \gls{JFT} is defined as follows: for $n=1,\dots,N$ and $i\geq 1$, 
\begin{align}
	\begin{aligned}\label{eq:JFT}
		\calF_{n,i}:L^2(\calJ)&\to\Real\\
		f&\mapsto \sum_{n'=1}^N \int_{\calT}f(n',\bt)\phi_n(n')\psi_{i}(\bt) \ud \tau(\bt),
	\end{aligned}
\end{align}
where $\phi_n(n')$ is the $n'$-th element of $\bphi_n$.
Using the \gls{JFT}, the signal is decomposed in the joint frequency domain indexed by $\set{(n, i)\given n=1\dots, N, i\geq1}$.

\section{Preliminaries: Random elements}\label[Appendix]{subsect:RE_intro}

In order to analyze the case where $f$ is a stochastic process indexed by $(v,\bt)$, we model $f$ as a random element \cite{JiaTay:J22, VakTarCho:87, HsiEub:15}. Consider a probability space $(\Omega,\calF,\mu)$, and a real separable Hilbert space $\calH$ with its norm-induced Borel $\sigma$-algebra $\calB$. A random element is defined as a measurable map $\bw:\Omega \mapsto \calH$, which induces a probability measure $\P$ on $(\calH,\calB)$ given by
\begin{align*}
	\P(B)=\mu(\bw^{-1}(B)),~\forall B\in \calB.
\end{align*}
Assume that $\E[\norm{\bw}]<\infty$. The mean of $\bw$ is defined as the element $m_{\bw}\in\calH$ such that 
\begin{align*}
	\angles{m_{\bw},\bh}
	=\E[\angles{\bw,\bh}], \forall \bh\in\calH.
\end{align*}
Assume that $\E[\norm{\bw}^2]<\infty$. The covariance of $\bw$ is defined as the operator $\bC_{\bw\bw}$ on $\calH$ such that 
\begin{align*}
	\angles{\bC_{\bw\bw}\bh, \bh'}
	&=\E[\angles{\bw-m_\bw,\bh}\angles{\bw-m_\bw,\bh'}], \forall \bh, \bh'\in\calH.
\end{align*}
In this paper we alternatively write $m_{\bw}$ and $\bC_{\bw\bw}$ as $\E[\bw]$ and $\Cov(\bw)$. It can be shown that $\Cov(\bw)$ is always compact, self-adjoint, positive semi-definite and trace-class. For a pair of random elements $(\bw_1, \bw_2):\Omega\rightarrow\calH_1\times\calH_2$ which satisfies $\E[\norm{(\bw_1, \bw_2)}^2]<\infty$, their cross-covariance operator is defined as the operator $\bC_{\bw_1\bw_2}:\calH_2\rightarrow\calH_1$ such that
\begin{align*}
	\angles{\bC_{\bw_1\bw_2}\bh_2,\bh_1}&=\E[\angles{\bw_1-\E[\bw_1],\bh_1}\angles{\bw_2-\E[\bw_2],\bh_2}], 
\end{align*}
for all $\bh_1\in\calH_1$, $\bh_2\in\calH_2$. We alternatively write $\bC_{\bw_1\bw_2}$ as $\Cov(\bw_1,\bw_2)$. The mean element, covariance operator and cross-covariance operator can be alternatively defined by Bochner integral \cite{HsiEub:15}. 

Let $\bh_1\in\calH_1$ and $\bh_2\in\calH_2$. We define $\bh_1\tens \bh_2$ as the following linear operator
\begin{align*}
	\bh_1\tens \bh_2:\calH_2&\to\calH_1\\
	\bh&\mapsto\angles{\bh,\bh_2}\bh_1.
\end{align*}
Note that $\bh_1\tens \bh_2$ is in the space of Hilbert-Schmidt operators from $\calH_2$ to $\calH_1$, which is a Hilbert space \cite[Theorem 4.4.5]{HsiEub:15}. Then $\bC_{\bw_1\bw_2}$ can be equivalently defined as $\E[(\bw_1 - \E[\bw_1])\tens(\bw_2- \E[\bw_2])]$.
The conditional expectation and covariance of a random element are defined as follows \cite[Section II.4.1]{VakTarCho:87}, \cite{KleSprSul:J20}: 

\begin{Definition}
	Suppose the random element $\bw$ takes values in a separable Hilbert space $\calH$, $\E[\norm{\bw}]<\infty$, and $\calF'$ is a sub $\sigma$-algebra of $\calF$. The conditional expectation of $\bw$ \gls{wrt} $\calF'$ is the random element $\bw_{\mathrm{cond}}\in\calF'$ such that $\E[\norm{\bw_{\mathrm{cond}}}]<\infty$ and
	\begin{align}\label{eq:def_cond_expc}
		\E[\bw_{\mathrm{cond}}I_{A}] = \E[\bw I_{A}],\ \forall A\in\calF',
	\end{align}
	where $I_A$ is the indicator function on the set $A$. We denote $\bw_{\mathrm{cond}}$ by $\E[\bw\given \calF']$. According to \cite[Proposition 4.1]{VakTarCho:87}, $\E[\bw\given \calF']$ always exists.
	
	The conditional covariance is defined as
	\begin{align*}
		&\Cov(\bw_1,\bw_2\given \calF') \\
		&= \E[(\bw_1-\E[\bw_1\given \calF'])\tens(\bw_2-\E[\bw_2\given \calF'])\given \calF']
	\end{align*}
	We write $\Cov(\bw,\bw\given \calF')$ as $\Cov(\bw\given \calF')$ for simplicity.
\end{Definition}

By the defining property \cref{eq:def_cond_expc} of conditional expectation it can be shown that
\begin{align}\label{eq:cond_exp_prop}
	\begin{aligned}
		\angles{\E[\bw\given\calF'], \bh} &= \E[\angles{\bw,\bh}\given\calF'],\\
		\angles{\E[\bw_1\tens\bw_2\given\calF'](\bh_2), \bh_1} &= \E[\angles{\bw_1,\bh_1}\angles{\bw_2,\bh_2}\given\calF'],
	\end{aligned}
\end{align} 
for all $\bh \in \calH$, $\bh_1\in\calH_1$, $\bh_2\in\calH_2$. From \cref{eq:cond_exp_prop} we know that $\E[\bw\given\calF']$ is uniquely defined. Let $\calF''$ be a sub $\sigma$-algebra of $\calF'$. Like random variables, the random elements also satisfy the property \cite[Section II.4.1]{VakTarCho:87}:
\begin{align*}
	\E[\E[\bw\given\calF']\given\calF''] = \E[\bw\given\calF''].
\end{align*}
Let $(\calI, \calF_\calI,\mu_\calI)$ be a $\sigma$-finite measure space. The stochastic process $\set{f(\omega, \bxi)\given\omega\in\Omega, \bxi\in\calI}$ can be modeled as a random element if it satisfies regularity conditions:

\begin{Theorem}\label{thm:Gauss_element}
	\cite[Theorem 2]{RajCam:J72}
	Suppose 
	\begin{enumerate}
		\item\label[condition]{thm:Gauss_element:mf} $f$ is a $\mu\times\mu_\calI$-measurable stochastic process.
		\item\label[condition]{thm:Gauss_element:L2} the paths of $f$ are in $L^2(\calI)$.
	\end{enumerate}
	Then the map
	\begin{align}\label{eq:map_random_elem}
		\begin{aligned}
			\Omega&\to L^2(\calI)\\
			\omega&\mapsto f(\omega,\cdot)
		\end{aligned}
	\end{align}
	is a random element with mean element $\E[f(\bxi)]\in L^2(\calI)$. Its covariance operator $\bC_f$ is the integral operator with kernel $\Cov(f(\bxi_1),f(\bxi_2))$. Specifically, if $f$ is \gls{GP}, then \cref{eq:map_random_elem} is a Gaussian random element, i.e., composing any linear functional with it will yield a Gaussian random variable.
\end{Theorem}
If we further assume that $\calI$ is a compact metric space and $\mu_\calI$ is a strictly positive Borel measure, and the function $\Cov(f(\bxi_1),f(\bxi_2))$ is continuous on $\calI\times\calI$, then it can be shown by Mercer's theorem \cite{SteSco:J12} that 
\begin{align}\label{eq:tr_is_int}
	\tr(\bC_f) = \int_{\calI} \Cov(f(\bxi_1),f(\bxi_2))\ud\mu_\calI.
\end{align}
In this paper we will make use of the following theorem which is more general than \cref{thm:Gauss_element}. The proof of it is included in \cref{proof:thm:cond_exp_cov} in the supplementary for completeness.
\begin{Theorem}\label{thm:cond_exp_cov}
	Suppose a stochastic process $f$ satisfies \cref{thm:Gauss_element:mf} and \cref{thm:Gauss_element:L2} in \cref{thm:Gauss_element}. $\calF'$ is a sub $\sigma$-algebra of the underlying probability space. Suppose $f$ and $\E[f(\bxi)\given\calF']\in L^2(\Omega\times\calI)$. Then
	\begin{align}
		\E[f\given\calF'] &= \E[f(\bxi)\given\calF'],\label{eq:cond_mean_func}\\
		\Cov(f\given\calF'):L^2(\calI)&\to L^2(\calI),\nn
		g(\cdot)&\mapsto\int_{\calI}\Cov(f(\bxi_1),f(\bxi_2)\given\calF')g(s)\ud \mu_\calI(\bxi_2).\label{eq:cond_cov_func}
	\end{align}
	If we further assume that $\calI$ is a compact metric space, and $\Cov(f(\bxi_1),f(\bxi_2)\given\calF')$ is continuous \gls{wrt} $(\bxi_1,\bxi_2)$, then we have
	\begin{align}\label{eq:cond_trace}
		\tr(\Cov(f\given\calF')) &= \E[\norm{f-\E[f\given\calF']}^2\given\calF']\nn
		&= \int_{\calI}\Var(f(\bxi)\given\calF')\ud \mu_\calI(\bxi).
	\end{align}
	In the above formulas, the \gls{LHS} are defined by moments of $f$ as a random element. The moments in \gls{RHS} are defined pointwise, as functions on $\calI$ or $\calI\times\calI$. 
\end{Theorem}

In this paper, the index set $\calI$ can be $\calV\times\calT$ or a subset of $\calV\times\calT$. We always assume that the conditions in \cref{thm:Gauss_element} are met for the stochastic processes in concern. In this case, we call the stochastic process $f$ as \gls{GRP} \cite{JiaTay:J22}. In statistical \gls{GSP}, a random graph signal is said to be \gls{WSS} if its covariance commutes with $\bA_G$ \cite{MarSegLeuRib:J17,PerVan:J17}. Analogously, in the \gls{GGSP} framework, a \gls{GRP} $f$ is said to be \gls{JWSS} if its covariance operator $\bC_f$ commutes with $\bS$ \cite{JiaTay:J22}.

\section{Preliminaries: KRR Reconstruction and Interpretation}\label[Appendix]{subsect:KRR_intro}

\gls{KRR} is a supervised learning approach that aims to learn a map from $\calX$ to $\calY$ where $\calY\subset\Real$. Given a set of training inputs and outputs, it searches for the best fitting function in a \gls{RKHS}. Given a symmetric positive semi-definite kernel
\begin{align*}
	k:\calX\times\calX&\to\calY\\
	(\bx,\bx')&\mapsto k(\bx,\bx'),
\end{align*}
the associated \gls{RKHS} $\calH_k$ is defined as the Hilbert space satisfying \cite[Definition 1]{BerTho:11}:
\begin{enumerate}
	\item $k(\cdot, \bx)\in\calH_k$ for all $\bx\in\calX$.
	\item $\angles{g,k(\cdot, \bx)}_{\calH_k} = g(\bx)$ for all $\bx\in\calX$ and $g\in\calH_k$.
\end{enumerate}
According to the Moore-Aronszajn theorem \cite[Theorem 3]{BerTho:11}, there exists a unique Hilbert space $\calH_k$ satisfying these conditions. When $\calX$ is a subset of Euclidean space, typical choices for $k$ include the polynomial kernel ($k(\bx,\bx') = (a\bx\T\bx'+1)^b$ with parameters $a\in\Real$, $b\in\bbN$), linear kernel (polynomial kernel with $a=1, b=1$), and \gls{RBF} kernel ($k(\bx,\bx')$ is a function of $\norm{\bx-\bx'}_\calX$).

Given a training set $\set{(\bx_m, y_m)\given \bx_m \in \calX, y_m\in\calY, m=1,\dots,M}$, \gls{KRR} searches for an optimal function in $\calH_k$ to fit the data by solving for 
\begin{align}\label{eq:KRR_prob}
	\hat{f} = \argmin_{\tilde{f}\in\calH_k} \sum_{m=1}^M \abs{\tilde{f}(\bx_m)-y_m}^2 + \mu J(\norm{\tilde{f}}_{\calH_k}),
\end{align}
where $J(\cdot)$ is an increasing function, and $\mu$ is a penalty weight. The representer theorem \cite[Theorem 4.2]{SchSmo:02} states that the optimal solution to \cref{eq:KRR_prob} takes the form 
\begin{align}\label{eq:KRR_optform}
	\hat{f} = \sum_{m=1}^M c_mk(\cdot,\bx_m),
\end{align}
where $c_m$, $m=1,\dots,M$, are coefficients to be determined.
By substituting \cref{eq:KRR_optform} into \cref{eq:KRR_prob}, the problem \cref{eq:KRR_prob} becomes an optimization over $\set{c_m}_{m=1}^M$. Specifically, when $J(\cdot) = (\cdot)^2$, problem \cref{eq:KRR_prob} is quadratic and its solution is given by
\begin{align}\label{eq:KRR_sol}
	(c_1,\dots,c_M)\T = (\bK+\mu\bI_M)^{-1}\by,
\end{align}
where $\bK=(k(\bx_i,\bx_j))_{i,j=1}^M \in\Real^{M\times M}$ and $\by=(y_1,\dots,y_M)\T$. In the sequel, we assume $J(\cdot) = (\cdot)^2$ unless otherwise stated. When $k$ is chosen as the linear kernel, \cref{eq:KRR_prob} is equivalent to learning a linear function from $\calX$ to $\calY$, i.e., linear regression. 

It is natural to consider whether we can recover any continuous function pointwise to within arbitrary fidelity with a sufficiently large number of samples by \gls{KRR}. This is achievable by employing a universal kernel $k$ \cite{MicYueHai:J06}. Let $\calX$ be a Hausdorff topological space and $\calZ\subset\calX$ be a compact subset. Let $\calC(\calZ)$ be the space of continuous functions on $\calZ$ with the supremum norm.  Define $\calK(\calZ):=\overline{\spn}\set{k(\cdot,\bx)\given \bx\in\calZ}$, where the closure is taken \gls{wrt} the norm in $\calC(\calZ)$. The kernel $k$ is said to be universal if $\calK(\calZ) = \calC(\calZ)$ for any compact $\calZ\subset\calX$. In other words, $\spn\set{k(\cdot,\bx)\given \bx\in\calZ}$ is dense in $\calC(\calZ)$.

Problem \cref{eq:KRR_prob} has a Bayesian interpretation. Consider a \gls{GP} $w$ with mean function zero and covariance function $k(\bx,\bx')$, denoted as $w\sim\GP(0,k)$. Given the noisy observations $y_m = w(\bx_m) + \epsilon_m$, $\epsilon_m\iid\dist{\calN}[0,\mu]$, the \gls{MAP} estimator of $w(\bx)$ is $\hat{f}(\bx)$ as defined in \cref{eq:KRR_optform} and \cref{eq:KRR_sol} for any $\bx\in\calX$.

The readers are referred to \cite{BerTho:11,HasTibFri:09} for more detailed discussions on \gls{RKHS} and \gls{KRR}.

\section{Proof of \cref{thm:limit_var}}\label[Appendix]{sect:proof:thm:limit_var}
In order to prove \cref{thm:limit_var}, we make use of the following lemmas. Their proofs are included in the supplementary for completeness.

\begin{Lemma}\label{lem:comp_trace_conv}
	Suppose a sequence of operators $\set{\bC_n}$ on a separable Hilbert space $\calH$, all of which are compact, self-adjoint, positive semi-definite and trace-class. Suppose $\bJ$ is a bounded linear operator from $\calH$ to $\calG$, where $\calG$ is also a separable Hilbert space. If $\lim\limits_{n\rightarrow\infty}\tr(\bC_n) = 0$, then $\lim\limits_{n\rightarrow\infty}\tr(\bJ\bC_n\bJ^*) = 0$.
\end{Lemma}	

\begin{Lemma}\label{lem:pullout}
	Suppose $\bw_1$ is a random element in $\calH_1$, and $\bw_2$ is a random element in $\calH_2$. $\calH_1$ and $\calH_2$ are separable Hilbert spaces. $\calF'$ is a sub $\sigma$-algebra of the underlying probability space. Suppose $\bw_2\in\calF'$, then we have
	\begin{align*}
		\E[\bw_1\circledast\bw_2\given \calF'] &= \E[\bw_1\given \calF'] \circledast \bw_2,\\
		\E[\bw_2\circledast\bw_1\given \calF'] &=  \bw_2\circledast \E[\bw_1\given \calF'].
	\end{align*}
\end{Lemma}

Using \cref{lem:pullout} we can simplify the definition of conditional covariance operator as
\begin{align*}
	\Cov(\bw_1,\bw_2\given \calF')
	= \E[\bw_1\circledast\bw_2\given\calF'] - \E[\bw_1\given\calF']\circledast\E[\bw_2\given\calF'].
\end{align*}

\begin{Lemma}\label{lem:EVVE}
	$\bC_{\bx_0|\by} 
	= \E[\Cov(\bx_0 \given \bz) \given \by(M_0)] + \Cov( \E[\bx_0\given \bz] \given \by(M_0))$.
\end{Lemma}

\begin{Lemma}\label{lem:cond_conv_0}
	Let $\bC_{\bz|\by}$ be the conditional covariance operator of $\bz$ given $\by(M_0)$. Then $\lim\limits_{M_0\rightarrow\infty}\tr(\bC_{\bz|\by}) = 0$ almost surely.
\end{Lemma}

\begin{Lemma}\label{lem:Gauss_element_mean_cov}
	The conditional expectation and covariance of $\bx_0$ given $\bz$ are as follows:
	\begin{align}\label{eq:Gauss_element_mean_cov}
		\begin{aligned}
			\E[\bx_0\given\bz] &= (\bC_{\bz\bz}^\dag\bC_{\bx_0\bz})^*\bz,\\
			\Cov(\bx_0\given\bz) &= \bC_{\bx_0} - \bC_{\bx_0\bz}\bC_{\bz\bz}^\dag\bC_{\bx_0\bz}^*,
		\end{aligned}
	\end{align} 
	where the operator $\bC_{\bz\bz}^\dag\bC_{\bx_0\bz}$ is bounded.
\end{Lemma}

\begin{proof}[Proof of \cref{thm:limit_var}]
	We can rewrite $\bC_{\bx_0 |\by}$ as follows:
	\begin{align}\label{eq:post_var_derive}
		\bC_{\bx_0 |\by}
		&= \E[\Cov(\bx_0 \given \bz) \given \by(M_0)] + \Cov(\E[\bx_0\given\bz]\given\by(M_0))\nn
		&= \Cov(\bx_0 \given \bz) + \Cov(\bC_{\bx_0\bz}\bC_{\bz\bz}^\dag\bz \given \by(M_0))\nn
		&=  \Cov(\bx_0 \given \bz) + \bC_{\bz\bz}^\dag\bC_{\bx_0\bz}\bC_{\bz|\by}(\bC_{\bz\bz}^\dag\bC_{\bx_0\bz})^*.
	\end{align}
	The first equality holds by \cref{lem:EVVE}. The second equality holds by the fact that $\Cov(\bx_0 \given \bz)$ is deterministic. 
	By taking trace and limit on \cref{eq:post_var_derive} we have 
	\begin{align*}
		&\lim_{M_0\rightarrow\infty} \tr(\bC_{\bx_0 |\by} - \Cov(\bx_0 \given \bz)) \\
		&= \lim_{M_0\rightarrow\infty} \tr(\bC_{\bz\bz}^\dag\bC_{\bx_0\bz}\bC_{\bz|\by}(\bC_{\bz\bz}^\dag\bC_{\bx_0\bz})^*).
	\end{align*}
	From \cref{lem:comp_trace_conv} and \cref{lem:cond_conv_0} we know that the \gls{RHS} tends to zero. By writing $\Cov(\bx_0 \given \bz)$ as \cref{eq:Gauss_element_mean_cov} and using \cref{eq:cond_trace} we conclude the proof.
\end{proof}

\section{Proof of \cref{thm:cond_exp_cov}}\label{proof:thm:cond_exp_cov}
\begin{proof}
	We first prove \cref{eq:cond_mean_func} and \cref{eq:cond_cov_func}. According to \cref{eq:cond_exp_prop}, it suffices to prove that for any $A\in\calF'$,
	\begin{align}
		&\E[\angles{\E[f(\bxi)\given\calF'], h(\bxi)}1_A] = \E[\angles{f, h}1_A],\label{eq:prove_mean_func}\\
		&\E[\angles{f(\bxi) - \E[f(\bxi)\given\calF'], h_1(\bxi)}\angles{f(\bxi) - \E[f(\bxi)\given\calF'], h_2(\bxi)}1_A] \nn
		&= \E[\angles{f-\E[f\given\calF'],h_1}\angles{f-\E[f\given\calF'],h_2}1_A].\label{eq:prove_cov_func}
	\end{align}
	We first prove \cref{eq:prove_mean_func} as follows:
	\begin{align*}
		\E[\angles{\E[f(\bxi)\given\calF'], h(\bxi)}1_A] &= \int_{A}\int_\calI \E[f(\bxi)\given\calF']h(\bxi) \ud \mu_\calI(\bxi) \ud\P\\
		&= \int_\calI\int_{A} \E[f(\bxi)\given\calF']h(\bxi) \ud\P\ud \mu_\calI(\bxi)\\
		&= \int_\calI \E[\E[f(\bxi)\given\calF']1_A]h(\bxi)\ud \mu_\calI(\bxi)\\
		&= \int_\calI\int_{A} f(\bxi) h(\bxi)\ud\P\ud \mu_\calI(\bxi)\\
		&= \int_{A}\int_\calI f(\bxi) h(\bxi)\ud \mu_\calI(\bxi)\ud\P\\
		&= \E[\angles{f,h}1_A].
	\end{align*}
	The integrals are exchangeable by Fubini's theorem.
	Similarly we prove \cref{eq:prove_cov_func} as follows:
	\begin{align*}
		&\E[\angles{f(\bxi) - \E[f(\bxi)\given\calF'], h_1(\bxi)}\angles{f(\bxi) - \E[f(\bxi)\given\calF'], h_2(\bxi)}1_A] \\
		&= \int_{A}\int_\calI\int_\calI (f(\bxi_1) - \E[f(\bxi_1)\given\calF'])h_1(\bxi_1)   (f(\bxi_2) - \E[f(\bxi_2)\given\calF'])h_2(\bxi_2)\ud \mu_\calI(\bxi_1) \ud \mu_\calI(\bxi_2) \ud\P\\
		&= \int_\calI\int_\calI \E[(f(\bxi_1) - \E[f(\bxi_1)\given\calF'])(f(\bxi_2) - \E[f(\bxi_2)\given\calF'])1_A]h_1(\bxi_1)h_2(\bxi_2)\ud \mu_\calI(\bxi_1) \ud \mu_\calI(\bxi_2)\\
		&= \int_{A}\int_\calI\int_\calI(f(\bxi_1) - \E[f(\bxi_1)\given\calF'])(f(\bxi_2) - \E[f(\bxi_2)\given\calF'])h_1(\bxi_1)h_2(\bxi_2)\ud \mu_\calI(\bxi_1) \ud \mu_\calI(\bxi_2)\\
		&= \E[\angles{f-\E[f\given\calF'],h_1}\angles{f-\E[f\given\calF'],h_2}1_A].
	\end{align*}
	To prove \cref{eq:cond_trace}, we first note that 
	\begin{align*}
		\E[\norm{f-\E[f\given\calF']}^21_A] &= \int_A \sum_{i=1}^{\infty} \angles{f-\E[f\given\calF'],e_i}^2\ud\P \\
		&= \sum_{i=1}^{\infty} \E [\angles{f-\E[f\given\calF'],e_i}^21_A] \\
		&= \sum_{i=1}^{\infty} \E[\angles{\Cov(f\given\calF')e_i, e_i}1_A]\\
		&= \E[\tr(\Cov(f\given\calF'))1_A].
	\end{align*}
	Thus, $\E[\norm{f-\E[f\given\calF']}^2\given\calF'] = \E[\tr(\Cov(f\given\calF'))\given\calF'] = \tr(\Cov(f\given\calF'))$.
	
	On the other hand, by Mercer's theorem \cite{SteSco:J12}, $\Cov(f(\bxi_1),f(\bxi_2)\given\calF')$ and $\Cov(f\given\calF')$ can be decomposed as 
	\begin{align*}
		\Cov(f(\bxi_1),f(\bxi_2)\given\calF') &= \sum_{i=1}^{\infty} \varrho_i\psi_i(\bxi_1)\psi_i(\bxi_2), \\
		\Cov(f\given\calF')(h) &= \sum_{i=1}^{\infty} \varrho_i\angles{\psi_i, h}\psi_i, \forall h\in L^2(\calI),
	\end{align*}
	where the convergence is uniform and absolute. $\varrho_i\geq0$. Besides, $\set{\psi_i}$ forms an orthonormal system in $L^2(\calI)$. Therefore,
	\begin{align*}
		\tr(\Cov(f\given\calF')) &= \sum_{i=1}^{\infty} \angles{\Cov(f\given\calF')\psi_i, \psi_i} \\
		&= \sum_{i=1}^{\infty} \varrho_i,
	\end{align*}
	and 
	\begin{align*}
		\int_{\calI} \Var(f(\bxi)\given\calF') &= \int_{\calI}\sum_{i=1}^{\infty}\varrho_i\psi_i(\bxi_1)\psi_i(\bxi_2)\mu_\calI(\bxi)\\
		&= \sum_{i=1}^{\infty}\varrho_i\int_{\calI}\psi_i(\bxi)\psi_i(\bxi)\mu_\calI(\bxi)\\
		&= \sum_{i=1}^{\infty}\varrho_i,
	\end{align*}
	which concludes the proof of \cref{eq:cond_trace}.
\end{proof}

\section{Proof of Lemmas for \cref{thm:limit_var}}\label{proof:thm:limit_var}
In this section, we prove the lemmas for the proof of \cref{thm:limit_var}.

\begin{proof}[proof of \cref{lem:comp_trace_conv}]
	Let $\set{\bh^{(n)}_i\given i=1,2,\dots}$ be the orthonormal basis of $\bC_n$. Let $\set{\lambda_i(\bC_n)\given i=1,2,\dots}$ be the corresponding eigenvalues. By definition of trace we have 
	\begin{align}\label{eq:trace_comp_sum}
		\tr(\bJ\bC_n\bJ^*) = \sum_{i=1}^\infty \angles{\bJ\bC_n\bJ^*\bh^{(n)}_i, \bh^{(n)}_i}
		= \sum_{i=1}^\infty\angles{\bC_n\bJ^*\bh^{(n)}_i, \bJ^*\bh^{(n)}_i}.
	\end{align}
	We now compute each term in \cref{eq:trace_comp_sum}. Assume that 
	\begin{align*}
		\bJ^*\bh^{(n)}_i = \sum_{j=1}^{\infty} \alpha_{ij}^{(n)} \bh^{(n)}_j.
	\end{align*}
	then we have 
	\begin{align*}
		\angles{\bC_n\bJ^*\bh^{(n)}_i, \bJ^*\bh^{(n)}_i} 
		&= \angles{\sum_{j=1}^{\infty} \alpha_{ij}^{(n)} \lambda_j(\bC_n) \bh^{(n)}_j, \sum_{j=1}^{\infty} \alpha_{ij}^{(n)} \bh^{(n)}_j}\\
		&= \sum_{j=1}^{\infty}(\alpha_{ij}^{(n)})^2\lambda_j(\bC_n).
	\end{align*}
	Substituting this result into \cref{eq:trace_comp_sum} we have 
	\begin{align*}
		\tr(\bJ\bC_n\bJ^*) 
		= \sum_{i=1}^\infty\sum_{j=1}^{\infty}(\alpha_{ij}^{(n)})^2\lambda_j(\bC_n)
		= \sum_{j=1}^{\infty}\lambda_j(\bC_n)\sum_{i=1}^\infty(\alpha_{ij}^{(n)})^2.
	\end{align*}
	Notice that $\bJ\bh^{(n)}_j = \sum\limits_{i=1}^\infty\alpha_{ij}^{(n)}\bh^{(n)}_i$, and $\norm{\bJ\bh^{(n)}_j}^2=\sum\limits_{i=1}^\infty(\alpha_{ij}^{(n)})^2\leq\norm{\bJ}^2$. Therefore,
	\begin{align*}
		\tr(\bJ\bC_n\bJ^*)\leq\norm{\bJ}^2\sum\limits_{j=1}^{\infty}\lambda_j(\bC_n) = \norm{\bJ}^2\tr(\bC_n) \rightarrow 0.
	\end{align*}
\end{proof}

\begin{proof}[proof of \cref{lem:pullout}]
	For any $\bh_1\in\calH_1$ and $\bh_2\in\calH_2$  we have
	\begin{align*}
		\angles{\E[\bw_1\circledast\bw_2\given \calF'](\bh_2),\bh_1} 
		&= \E[\angles{\bw_2, \bh_2}\angles{\bw_1, \bh_1}\given \calF']\\
		&= \angles{\bw_2, \bh_2}\E[\angles{\bw_1, \bh_1}\given \calF']\\
		&= \angles{\bw_2, \bh_2}\angles{\E[\bw_1\given\calF'], \bh_1}.
	\end{align*}
	The first and third equality are obtained by \cref{eq:cond_exp_prop}. The second equality is due to the fact that $\bw_2\in\calF'$. On the other hand, by definition we have 
	\begin{align*}
		\angles{\E[\bw_1\given \calF'] \circledast \bw_2(\bh_2), \bh_1} = \angles{\bw_2, \bh_2}\angles{\E[\bw_1\given\calF'], \bh_1},
	\end{align*}
	which concludes the proof of the first equation. The second equation can be proved by a similar argument.
\end{proof}

%
%
\begin{proof}[proof of \cref{lem:EVVE}]
	To prove this equality, we mainly make use of the fact that $\by(M_0)\in\sigma(\bz,\set{\epsilon_m}, \calS(M_0))$. We write $\sigma(\bz,\set{\epsilon_m}, \calS(M_0))$ as $\calF_0$ for simplicity. Notice that $\Cov(\bx_0 \given \bz) = \Cov(\bx_0 \given \calF_0)$ and $\E[\bx_0\given \bz] = \E[\bx_0\given \calF_0]$ since both $\set{\epsilon_m}$ and $\calS(M_0)$ are jointly independent of the \gls{GP} $f$. The first term in the \gls{RHS} of \cref{lem:EVVE} can be computed as follows:
	
	\begin{align}\label{eq:EV_1}
		\E[\Cov(\bx_0 \given \calF_0) \given \by(M_0)] 
		&= \E[\E[(\bx_0 - \E[\bx_0\given\calF_0])\circledast(\bx_0 - \E[\bx_0\given\calF_0])\given \calF_0]\given\by(M_0)]\nn
		&= \E[(\bx_0 - \E[\bx_0\given\calF_0])\circledast(\bx_0 - \E[\bx_0\given\calF_0])\given\by(M_0)]\nn
		&= \E[\bx_0\circledast\bx_0\given\by(M_0)] - \E[\bx_0\circledast\E[\bx_0\given\calF_0]\given\by(M_0)] \nn
		&- \E[\E[\bx_0\given\calF_0]\circledast\bx_0\given\by(M_0)]+ \E[\E[\bx_0\given\calF_0]\circledast\E[\bx_0\given\calF_0]\given\by(M_0)].
	\end{align}
	
	The second equality is derived by the fact that $\by(M_0)\in\calF_0$. We further use this fact and \cref{lem:pullout} to calculate the second and third term in \cref{eq:EV_1}:
	
	\begin{align*}
		\E[\bx_0\circledast\E[\bx_0\given\calF_0]\given\by(M_0)]&= \E[\E[\bx_0\circledast\E[\bx_0\given\calF_0]]\given\calF_0]\given\by(M_0)]\\
		&= \E[\E[\bx_0\given\calF_0]\circledast\E[\bx_0\given\calF_0]\given\by(M_0)],\\
		\E[\E[\bx_0\given\calF_0]\circledast\bx_0\given\by(M_0)]&= \E[\E[\E[\bx_0\given\calF_0]\circledast\bx_0]\given\calF_0]\given\by(M_0)]\\
		&= \E[\E[\bx_0\given\calF_0]\circledast\E[\bx_0\given\calF_0]\given\by(M_0)].\\
	\end{align*}
	
	Substituting this result into \cref{eq:EV_1}, we obtain
	
	\begin{align}\label{eq:EV_2}
		\E[\Cov(\bx_0 \given \calF_0) \given \calF_0) \given \by(M_0)] = \E[\bx_0\circledast\bx_0\given\by(M_0)] - \E[\E[\bx_0\given\calF_0]\circledast\E[\bx_0\given\calF_0]\given\by(M_0)].
	\end{align}
	
	Using a similar argument as above, we have
	
	\begin{align}\label{eq:VE}
		\Cov(\E[\bx_0\given \calF_0] \given \by(M_0)) = \E[\E[\bx_0\given\calF_0]\circledast\E[\bx_0\given\calF_0]\given\by(M_0)] - \E[\bx_0\given\by(M_0)]\circledast\E[\bx_0\given\by(M_0)].
	\end{align}
	
	By combining \cref{eq:EV_2} and \cref{eq:VE}, we obtain the conclusion in \cref{lem:EVVE}.
\end{proof}
\begin{proof}[proof of \cref{lem:cond_conv_0}]
	According to \cite[Theorem 3]{KoePfa:J21},
	\begin{align*}
		\sup_{(v,\bt)\in\calJ_S} \Var(f(v,\bt) \given \by(M_0)) \rightarrow 0
	\end{align*}
	almost surely monotonically, hence 
	
	\begin{align*}
		\tr(\bC_{\bz|\by}) = \int_{\calJ_S} \Var(f(v,\bt) \given \by(M_0)) \ud \zeta(v,\bt)\rightarrow0,\as.
	\end{align*}
	
\end{proof}

\begin{proof}[proof of \cref{lem:Gauss_element_mean_cov}]
	We first prove that $(\bx_0,\bz)$ meets the compatible condition in \cite[Section 4.2]{KleSprSul:J20}, i.e., $\ima(\bC_{\bz \bx_0})\subset\ima(\bC_{\bz\bz})$. From \cref{eq:cov_operator_forms} and the fact that $\bK_{G}$ is full-rank we know that $\ima(\bC_{\bz \bx_0}) = \spn{\set{\bk_{G,0*}}}\otimes\ima(\bH_0)$ and $\ima(\bC_{\bz \bz}) = \Real^{T-1}\otimes\ima(\bH)$. Hence it suffices to prove that $\ima(\bH_0)\in\ima(\bH)$, which can be shown by definition of $\bH$ and $\bH_0$.
	
	Second, since $f$ is a \gls{GP}, according to \cref{thm:Gauss_element}, $(\bx_0,\bz)$ is a Gaussian random element on $(L^2(\calJ_S\cup(\set{v_0}\times \calT_0)),\calB)$. Then we obtain \cref{eq:Gauss_element_mean_cov} by using \cite[Theorem 4.8]{KleSprSul:J20} and \cite[Section 6]{KleSprSul:J20}.
\end{proof}

\section{proof of \cref{thm:var_bound}}\label{proof:thm:var_bound}
\begin{proof}
	We prove this theorem in two steps: first, we prove that with large probability there are enough sample points in a small neighborhood of $\calQ$. Then, we prove that since the neighborhood is small, we can asymptotically upper bound $\Var(f(v_0,\bt_0) \given \by(M_0))$ by $\Var(f(v_0,\bt_0) \given f(\calQ))$.
	
	Consider the neighborhood of $\bt_0$: $B(\bt_0,\delta)\subset\calT$. Let $C_D = \dfrac{\tau(B(\bt_0,1))}{\tau(\calT)}$, then $\tau(B(\bt_0,\delta)) = \delta^D\tau(\calT)C_D$. Let $\calS(v,M_0,\delta) = \calS(v;M_0)\bigcap(\set{v}\times B(\bt_0,\delta))$ be the sample points that falls into $B(\bt_0,\delta)$ on vertex $v$. Then on each vertex $v$, $\abs{\calS(v,M_0,\delta)}$ is a Binomial random variable $n_0\iid\text{Binom}(M_0, C_D\delta^D)$. By Chebyshev's inequality we have
	\begin{align*}
		\P(\abs{n_0-M_0C_D\delta^D}\geq M_0C_D\delta^D(1-c_0))\leq \frac{1-C_D\delta^D}{(1-c_0)^2M_0C_D\delta^D}.
	\end{align*}
	Besides, due to the symmetry of the binomial distribution, we have
	\begin{align*}
		\P(n_0\leq c_0M_0C_D\delta^D)
		&=\P(n_0-M_0C_D\delta^D\leq-M_0C_D\delta^D(1-c_0))\\
		&= \ofrac{2}\P(\abs{n_0-M_0C_D\delta^D}\geq M_0C_D\delta^D(1-c_0)).
	\end{align*}
	Therefore, the number of samples in $B(\bt_0,\delta)$ can be lower bounded by
	\begin{align*}
		\P(n_0> c_0M_0C_D\delta^D)\geq 1-\ofrac{2}\frac{1-C_D\delta^D}{(1-c_0)^2M_0C_D\delta^D}.
	\end{align*}
	For ease of notation, we use $m_0$ to denote $c_0M_0C_D\delta^D$ in the proof. Since the samples are obtained independently on each vertex, the probability that every vertex $v$ in $\calN_{d}$ has more than $m_0$ sampled instances in $B(\bt_0,\delta)$ can be lower bounded by
	\begin{align}\label{eq:prob_lowerbound_1}
		\P\parens*{\abs{\calS(v,M_0,\delta)}>m_0,\forall v\in\calN_d}=
		\parens*{\P(n_0>m_0)}^{N_d}\geq \parens*{1-\ofrac{2}\frac{1-C_D\delta^D}{(1-c_0)^2M_0C_D\delta^D}}^{N_d}.
	\end{align}
	In the sequel, we will work on this event. We will prove that with more than $m_0$ samples in $B(\bt_0,\delta)$ on each $v\in\calN_d$, we are able to upper bound $\Var(f(v_0,\bt_0) \given \by(M_0))$ by $\Var(f(v_0,\bt_0) \given f(\calQ))$.
	Let $\calS(v,M_0,\delta,m_0)\subset\calS(v,M_0,\delta)$ be any subset with cardinality $m_0$. Let $\calS(v,M_0,\delta,m_0)=\set{v}\times \bt^{(v)}$, where $\bt^{(v)}=\set{\bt^{(v)}_1,\dots,\bt^{(v)}_{m_0}}$. We write $\bY(M_0,\delta):=(y(v,\bt^{(v)}_j))_{v\in\calN_{d}(v_0),j\in[m_0]}\in\Real^{N_d\times m_0}$.
	According to \cite[Lemma 9]{KoePfa:J21}, the posterior variance can be bounded by
	\begin{align}\label{eq:cond_less_data}
		\Var(f(v_0,\bt_0) \given \by(M_0)) 
		&\leq
		\Var(f(v_0,\bt_0) \given \bY(M_0,\delta))\nn
		&=k_\calT(\bt_0,\bt_0)k_G(v_0,v_0) - \bk\T\bB^{-1}\bk,
	\end{align}
	where $\bk$ and $\bB$ are calculated as in \cref{eq:GP_post_var}:
	\begin{align*}
		\bk_\calT(\bt_0,\bt^{(v)})&=(k_\calT(\bt_0,\bt_1^{(v)}),\dots,k_\calT(\bt_0,\bt_{m_0}^{(v)}))\T\in\Real^{m_0},\\
		\bK_\calT(\bt^{(u)},\bt^{(v)}) &= (k_\calT(\bt_{i}^{(u)},\bt_{j}^{(v)}))_{i,j\in[m_0]}\in\Real^{m_0\times m_0},
	\end{align*}
	\begin{align*}
		\bk&=(k_G(v_0,v)\bk_\calT(\bt_0,\bt^{(v)})\T)_{v\in\calN_{d}(v_0)}\T\in\Real^{N_dm_0},\\
		\bB &= (k_G(u,v)\bK_\calT(\bt^{(u)},\bt^{(v)}))_{u,v\in\calN_{d}(v_0)}+ \sigma^2\bI_{N_dm_0}\in\Real^{N_dm_0\times N_dm_0}.
	\end{align*}
	Intuitively, when $\delta$ is small enough, the points in $\bt^{(v)}$ will be close to $\bt_0$, thus all $\bt^{(v)}_i$ in $\bk$ and $\bB$ can be replaced by $\bt_0$. Following this idea, we define
	\begin{align*}
		\bkappa &= k_\calT(\bt_0,\bt_0)\bk_G(v_0,\calN_d)\otimes\bone_{m_0}\\
		\bB' &= k_\calT(\bt_0,\bt_0)\bK_G(\calN_d,\calN_d)\otimes\bone_{m_0}\bone_{m_0}\T + \sigma^2\bI_{N_dm_0}.
	\end{align*}
	We aim to approximate \cref{eq:cond_less_data} by replacing $\bk$ with $\bkappa$ and $\bB$ with $\bB'$.
	\begin{align}\label{eq:bound_appro_1}
		&\abs{\Var(f(v_0,\bt_0) \given \bY(M_0,\delta))-k_\calT(\bt_0,\bt_0)l_1}\nn
		&= \abs{k_\calT(\bt_0,\bt_0)\bk_G(v_0,\calN_d)\T\bK_G(\calN_d,\calN_d)^{-1}\bk_G(v_0,\calN_d)-\bk\T\bB^{-1}\bk}\nn
		&\leq \abs{k_\calT(\bt_0,\bt_0)\bk_G(v_0,\calN_d)\T\bK_G(\calN_d,\calN_d)^{-1}\bk_G(v_0,\calN_d)-\bkappa\T\bB'^{-1}\bkappa}\nn
		&+\abs{\bkappa\T\bB'^{-1}\bkappa-\bk\T\bB^{-1}\bk}
	\end{align}
	we are going to treat the two terms in \cref{eq:bound_appro_1} respectively. We denote the first term as $(i)$ and the second term as $(ii)$. To this end, we first need to calculate $\bB'^{-1}$. By respectively calculating the eigenvalues and eigenvectors of $\bB'$ on $\Real^{N_d}\otimes\spn\{\bone_{m_0}\}$ and $\Real^{N_d}\otimes\spn\{\bone_{m_0}\}^{\perp}$, it can be shown that
	\begin{align*}
		\bB'^{-1}=(k_\calT(\bt_0,\bt_0)m_0\bK_G(\calN_d,\calN_d)+\sigma^2\bI_{N_d})^{-1}\otimes\ofrac{m_0}\bone_{m_0}\bone_{m_0}\T + \ofrac{\sigma^2}\bI_{N_d}\otimes(\bI_{m_0}-\ofrac{m_0}\bone_{m_0}\bone_{m_0}\T).
	\end{align*}
	To simplify the notation, we define the matrix
	\begin{align*}
		\bQ_G=(k_\calT(\bt_0,\bt_0)m_0\bK_G(\calN_d,\calN_d)+\sigma^2\bI_{N_d})^{-1},
	\end{align*}
	so that 
	\begin{align*}
		\bkappa\T\bB'^{-1}\bkappa = k_\calT(\bt_0,\bt_0)^2m_0\bk_G(v_0,\calN_d)\T\bQ_G\bk_G(v_0,\calN_d).
	\end{align*}
	\begin{align}\label{eq:bound_i_1}
		(i) &= \abs{\bk_G(v_0,\calN_d)\T(k_\calT(\bt_0,\bt_0)\bK_G(\calN_d,\calN_d)^{-1}-k_\calT(\bt_0,\bt_0)^2m_0\bQ_G)\bk_G(v_0,\calN_d)}\nn
		&\leq\abs{k_\calT(\bt_0,\bt_0)}\norm{\bk_G(v_0,\calN_d)}_2^2\norm{\bK_G(\calN_d,\calN_d)^{-1}-k_\calT(\bt_0,\bt_0)m_0\bQ_G}_2.
	\end{align}
	Notice that $\bK_G(\calN_d,\calN_d)$ and $\bQ_G$ has the same set of eigenvectors. Specifically, if $\bpsi$ is $\bK_G(\calN_d,\calN_d)$'s eigenvector associated with eigenvalue $\alpha$, then it is $\bQ_G$'s eigenvector associated with eigenvalue $(k_\calT(\bt_0,\bt_0)m_0\alpha+\sigma^2)^{-1}$. Let $\sigma_{\min}>0$ be the minimum eigenvalue of $\bK_G(\calN_d,\calN_d)$. Using this relationship we can derive a bound for the norm of the matrix difference
	\begin{align*}
		\norm{\bK_G(\calN_d,\calN_d)^{-1}-k_\calT(\bt_0,\bt_0)m_0\bQ_G}_2
		&=\ofrac{\sigma_{\min}}-\ofrac{\sigma_{\min}+\sigma^2/(k_\calT(\bt_0,\bt_0)m_0)}\\
		&< \ofrac{\sigma_{\min}^2}\frac{\sigma^2}{k_\calT(\bt_0,\bt_0)m_0}.
	\end{align*}
	By substituting this result into \cref{eq:bound_i_1}, and noticing that the $k_\calT(\bt_0,\bt_0)$ and $\bk_G(v_0,\calN_{d})$ do not depend on $m_0$, we have
	\begin{align}\label{eq:bound_i_2}
		(i)<\frac{C_1}{m_0},
	\end{align}
	where $C_1$ is a constant which only depends on $d$.
	
	We find the upper bound for $(ii)$ by triangle inequality:
	\begin{align}\label{eq:bound_ii_1}
		(ii)&\leq \abs{\bkappa\T\bB'^{-1}\bkappa-\bkappa\T\bB^{-1}\bkappa} + \abs{\bkappa\T\bB^{-1}\bkappa-\bk\T\bB^{-1}\bkappa} + \abs{\bk\T\bB^{-1}\bkappa-\bk\T\bB^{-1}\bk}\nn
		&\leq\norm{\bkappa}_2^2\norm{\bB'^{-1}-\bB^{-1}}_2+\norm{\bkappa-\bk}_2\norm{\bB^{-1}}_2\norm{\bkappa}_2+\norm{\bkappa-\bk}_2\norm{\bB^{-1}}_2\norm{\bk}_2.
	\end{align}
	Then it suffices to find bounds for the norms of vectors and matrices in \cref{eq:bound_ii_1}. By definition, we have
	\begin{align*}
		\norm{\bkappa}_2 =  k_\calT(\bt_0,\bt_0)\norm{\bk_G(v_0,\calN_{d})}_2m_0^{\ofrac{2}}.
	\end{align*}
	Since $k_\calT$ is continuous and $\calT$ is compact, $k_\calT$ can achieve its maximum, denoted as $M(k_\calT):=\max\limits_{(\bs,\bt)}\{k_\calT(\bs,\bt)\}$. Besides, $k_\calT$ is Lipschitz continuous with Lipschitz constant $L(k_\calT)$. Then we have
	\begin{align*}
		\norm{\bk}_2&\leq M(k_\calT) \norm{\bk_G(v_0,\calN_{d})}_2m_0^{\ofrac{2}},\\
		\norm{\bkappa-\bk}_2&\leq \norm{\bk_G(v_0,\calN_{d})}_2L(k_\calT)\delta m_0^{\ofrac{2}}\\
		\norm{\bB-\bB'}_2&\leq \norm{\bB-\bB'}_\infty
		\leq\sqrt{2}\norm{\bk_G(v_0,\calN_d)}_1L(k_\calT)\delta m_0.
	\end{align*}
	By definition we know that $\bB\succeq\sigma^2\bI_{N_d^1m_0}$, so $\bB^{-1}\preceq\ofrac{\sigma^2}\bI_{N_d^1m_0}$, $\norm{\bB^{-1}}_2\leq\ofrac{\sigma^2}$. Using the same argument we have $\norm{\bB'^{-1}}\leq\ofrac{\sigma^2}$.
	\begin{align*}
		\norm{\bB'^{-1}-\bB^{-1}}_2 
		&= \norm{\bB^{-1}(\bB-\bB')\bB'^{-1}}_2\\
		&= \norm{\bB^{-1}}_2\norm{\bB-\bB'}_2\norm{\bB'^{-1}}_2\\
		&\leq \sqrt{2}\sigma^{-4}\norm{\bk_G(v_0,\calN_d)}_1L(k_\calT)\delta m_0.
	\end{align*}
	Combining all the bounds on vectors and matrices' norms with \cref{eq:bound_ii_1} we obtain that
	\begin{align}\label{eq:bound_ii_2}
		(ii)\leq C_{2}\delta m_0^2+C_{3}\delta m_0,
	\end{align}
	where $C_2$ and $C_3$ are constants only depend on $d$. By combining \cref{eq:bound_i_2}, \cref{eq:bound_ii_2} with \cref{eq:bound_appro_1} we obtain that
	\begin{align*}
		\abs{\Var(f(v_0,\bt_0) \given \bY(M_0,\delta))-k_\calT(\bt_0,\bt_0)l_1}\leq \frac{C_1}{m_0} + C_{2}\delta m_0^2+C_{3}\delta m_0.
	\end{align*}
	Now we are to examine the asymptotic case when $M_0\rightarrow\infty$. Recall that $m_0=c_0M_0C_D\delta^D$. If we let $\delta=M_0^{-\beta}$ where $\beta>0$, we have
	\begin{align}\label{eq:bound_appro_2}
		\abs{\Var(f(v_0,\bt_0) \given \bY(M_0,\delta))-k_\calT(\bt_0,\bt_0)l_1}\leq C_1c_0^{-1}M_0^{\beta D-1} + C_{2}c_0^2M_0^{2-(2D+1)\beta}+C_{3}c_0M_0^{1-(D+1)\beta},
	\end{align}
	where the constant $C_D$ is absorbed by $C_1,C_2$ and $C_3$. By requiring the powers of $M_0$ to be negative, $\beta$ should be in the range $(\frac{2}{2D+1},\frac{1}{D})$. By adjusting $\beta$ in this range, the best rate is achieved when $\hat{\beta}=\frac{3}{3D+1}$. By substituting this into \cref{eq:bound_appro_2}, we obtain
	\begin{align}\label{eq:bound_appro_3}
		\abs{\Var(f(v_0,\bt_0) \given \bY(M_0,\delta))-k_\calT(\bt_0,\bt_0)l_1}\leq (C_1c_0^{-1}+ C_{2}c_0^2)M_0^{-\ofrac{3D+1}} + C_3c_0M_0^{-\frac{2}{3D+1}}.
	\end{align}
	On the other hand, by substituting $\delta=M_0^{-\hat{\beta}}$ into \cref{eq:prob_lowerbound_1}, we have
	\begin{align}\label{eq:prob_lowerbound_2}
		\P\parens*{\abs{\calS(v,M_0,\delta)}>m_0,\forall v\in\calN_d(v_0)}&\geq \parens*{1-\ofrac{2}\frac{1-C_D\delta^D}{(1-c_0)^2M_0C_D\delta^D}}^{N_d}\nn
		&\geq
		\parens*{1-\ofrac{2}\frac{1}{(1-c_0)^2C_DM_0^{\ofrac{3D+1}}}}^{N_d}.
	\end{align}
	Finally, by combining \cref{eq:cond_less_data}, \cref{eq:bound_appro_3} and \cref{eq:prob_lowerbound_2}, we conclude the proof.
\end{proof}

\bibliographystyle{IEEEtran}
\bibliography{IEEEabrv,StringDefinitions,refs}

\end{document}